\documentstyle[aasms]{article}

\def\lae{\mathrel{<\kern-1.0em\lower0.9ex\hbox{$\sim$}}}
\def\gae{\mathrel{>\kern-1.0em\lower0.9ex\hbox{$\sim$}}}

\font\fsmall=cmr10

\def\fdg{\hbox{$.\!\!^\circ$}}
\def\farcm{\hbox{$.\mkern-4mu^\prime$}}
\def\farcs{\hbox{$.\!\!^{\prime\prime}$}}

\articleid{11}{14}

\slugcomment{To appear in the Astronomical Journal}

\begin{document}

\title{Spectroscopic Binaries in Globular Clusters. II. A Search for Long-Period Binaries in M22}

\author{Patrick C\^ot\'e\altaffilmark{1,2}}
\affil{ Dominion Astrophysical Observatory, Herzberg Institute of Astrophysics,
    National Research Council of Canada, 5071 W. Saanich Road, R.R.5, Victoria, BC, V8X 4M6, Canada \\
    cote@dao.nrc.ca}

\author{Carlton Pryor\altaffilmark{3}}
\affil{Department of Physics and Astronomy, Rutgers, The State University, Box 0849 \\ Piscataway, 
New Jersey 08855-0849, USA \\ pryor@physics.rutgers.edu}

\and

\author{Robert D. McClure\altaffilmark{3}, J. M. Fletcher\altaffilmark{3}, James E. Hesser\altaffilmark{3}}
\affil{ Dominion Astrophysical Observatory, Herzberg Institute of Astrophysics,
    National Research Council of Canada, 5071 W. Saanich Road, R.R.5, Victoria, BC, V8X 4M6, Canada \\
    mcclure@dao.nrc.ca,fletcher@dao.nrc.ca,hesser@dao.nrc.ca}

% Notice that each of these authors has alternate affiliations, which
% are identified by the \altaffilmark after each name.  The actual alternate
% affiliation information is typeset in footnotes at the bottom of the
% first page, and the text itself is specified in \altaffiltext commands.
% There is a separate \altaffiltext for each alternate affiliation
% indicated above.

\altaffiltext{1}{Visiting Astronomer, Cerro Tololo Inter-American Observatory. 
CTIO is operated by AURA, Inc.\ under contract to the National Science
Foundation.} 

\altaffiltext{2}{Visiting Astronomer, Kitt Peak National Observatory. 
KPNO is operated by AURA, Inc.\ under contract to the National Science
Foundation.} 

\altaffiltext{3}{Visiting Astronomer, Canada-France-Hawaii Telescope operated
by the National Research Council of Canada, the Centre National de la Recherche
Scientifique de France and the University of Hawaii.}

% The abstract environment prints out the receipt and acceptance dates
% if they are relevant for the journal style.  For the aasms style, they
% will print out as horizontal rules for the editorial staff to type
% on, so long as the author does not include \received and \accepted
% commands.  This should not be done, since \received and \accepted dates
% are not known to the author.

\begin{abstract}

A catalog of 383 radial velocities (median accuracy $\simeq$ 1 km s$^{-1}$) for red giants in the
Galactic globular cluster M22 has been compiled from the literature and from new
observations accumulated between 1972 and 1994.
This 22-year baseline is the longest available for any sample of globular cluster stars.
Using 333 repeat velocities for 109 cluster members,
we have carried out a search for spectroscopic binaries with periods in the range
0.2 $\lae$ P $\le$ 40 years and with mass ratios between 0.1 and 1.0. 
Although the radial velocities for these evolved stars show clear evidence for an atmospheric ``jitter"
whose magnitude depends on luminosity, no star is convincingly found
to exhibit a velocity variation greater than 7 km s$^{-1}$. By comparing
the observed velocity variations to those found in a series of Monte-Carlo
simulations, we estimate the cluster binary fraction to be $x_b$ = 0.01$^{+0.10}_{-0.01}$ (circular orbits)
and $x_b$ = 0.03$^{+0.16}_{-0.03}$ (thermal orbits), where the uncertainties are 90\% confidence limits.
These results are to be compared to the corresponding binary fraction of $x_b$ = 0.12$\pm$0.03 for
nearby, solar-type stars having similar mass ratios and periods.
We speculate that both the relative abundances of short- and long-period binaries
in globular clusters and the large differences in measured binary fractions for clusters with high
binary ionization rates (M22, $\omega$ Cen) compared to those clusters with
low ionization rates (M71, M4, NGC 3201) 
point to a frequency-period distribution in which ``soft" binaries have been disrupted by stellar encounters.
Finally, we note that none of the three CH stars in our survey shows evidence for velocity
variations; this is in stark contrast to {\sl field} CH stars, virtually all of which are 
binaries. We argue that binaries in M22 which have binding energies similar to field CH stars are unlikely
to have been disrupted
and suggest that the cluster CH stars are otherwise normal red giants which lie in the carbon-enriched 
tail of the cluster metallicity distribution function.
\end{abstract}

\keywords{globular clusters: individual (M22)  --- stars: binaries --- stars: carbon --- techniques: radial velocities}

% That's it for the front matter.  On to the main body of the paper.
% We'll only put in tutorial remarks at the beginning of each section
% so you can see entire sections together.
%
% In the first two sections, you should notice the use of the LaTeX \cite
% command to identify citations.  The citations are tied to the
% reference list via symbolic tags.  We have chosen the first three
% characters of the first author's name plus the last two numeral of the
% year of publication.  The corresponding reference has a \bibitem
% command in the reference list below.
%
% Please go to the LaTeX manual for a complete description of the
% \cite-\bibitem mechanism.

\section{Introduction}

The binary fraction, $x_b$, is a fundamental parameter of any stellar population. For dense
stellar systems like globular clusters, it is also an essential ingredient for realistic models of their
dynamical evolution. The first efforts to measure $x_b$ for globular clusters led to the surprising 
conclusion that they were markedly deficient in binaries relative to the field (e.g., Gunn \& Griffin 1979).
This result became difficult to understand in the light of numerical simulations (e.g., Cohn 1980) 
which demonstrated repeatedly that, in the absence of an additional heat source such as that provided by 
a population of primordial binaries,
most Galactic globular clusters should by now have settled into a state of core collapse.
Early ground-based surveys, however, indicated that, at most, 
only $\sim$ 20\% of Galactic globular clusters exhibit the telltale central brightness cusp
predicted by core-collapse models (Djorgovski \& King 1986).
The large numbers of blue stragglers seen in many globulars is additional indirect evidence 
for the existence of cluster binaries. Although it remains to be seen whether the
bulk of these cluster blue stragglers formed via stellar collisions (e.g., Hoffer 1983) or through 
coalescence of close pairs (Zinn \& Searle 1976), it now appears likely that binaries
play a key role in their formation.

A number of recent observational studies have concluded that the globular cluster binary fraction is roughly comparable
to that of the Population I field (e.g., Hut et al. 1992; C\^ot\'e et al. 1994; Yan \& Mateo 1994). Nevertheless,
our knowledge of the {\sl shape} of the distribution of binary star periods in globular
clusters remains remarkably limited. For instance, the distribution of orbital periods among
nearby, solar-type stars is closely Gaussian in $\log {\rm P}$ with a peak near 180 yr and 
a dispersion $\sigma$$_{\rm log~P}$ $\simeq$ 2.3 (Duquennoy \& Mayor 1991). Does the 
globular cluster period distribution differ significantly from this field distribution? 
There are strong reasons to believe that it {\sl must}, since stellar
encounters will tend to disrupt ``soft" binaries and harden already ``hard" systems (Heggie 1975).
Hills (1984) found that the semi-major axis of the widest binary which is 
expected to have escaped disruption by encounters with single stars is
$${\rm a_c} = 12.4~{\rm AU}\Biggl({{\rm M_1} + {\rm M_2} \over 1.4{\rm M}_{\odot}}\Biggr)\Biggl({10~{\rm km~s}^{-1} 
\over \sigma}\Biggr)^2,\eqno{(1)}$$
where $\sigma$ is the three-dimensional rms cluster velocity dispersion in km s$^{-1}$ and M$_1$ and 
M$_2$ are the primary and secondary masses in solar units.  
Binaries with separations greater than a$_{\rm c}$ are expected to be disrupted in a Hubble time, provided
$\sigma$ and the cluster number density of stars, $n$, satisfy (Pryor et al. 1996)
$$\Biggl( {n \over {1~{\rm pc}^{-3}}}\Biggr)\Biggl({5~{\rm km~s}^{-1} \over {{\sigma}}}\Biggr)^{3} > 106.\eqno{(2)}$$
This relation, which is based on the ionization rate of Hut \& Bahcall (1983) and assumes 0.8M$_{\odot}$ stars,
is satisfied at the centers of most clusters.
According to Pryor \& Meylan (1993), Galactic globular clusters
have 2 $\lae \sigma \lae 25$ km s$^{-1}$ with a mean near $\simeq$ 8 km s$^{-1}$.
Consequently, the widest surviving systems will have
$3 \lae {\rm P_c} \lae 5000$ years, with a typical upper period cutoff of P$_{\rm c}$ $\simeq$ 80 years
(assuming M$_1$ and M$_2$ = 0.8M$_{\odot}$).
Dynamical processes are therefore likely to have played an important role in modifying 
the initial distribution of orbital periods in globular clusters.

Unfortunately, existing data are not well suited to the task of detecting long-period binaries. 
For instance, studies of binaries based on the so-called ``second sequence" in the cluster color-magnitude diagram
are {\sl period degenerate} (i.e., equally sensitive to binaries of {\sl all} periods). 
On the other hand,
searches for eclipsing binaries (e.g., Mateo 1996)
are sensitive only to systems with periods near one day and are 
virtually blind to binaries with periods in excess of $\sim$ 10 days (Mateo 1993). 
Although the situation is greatly improved for radial velocity surveys,
studies to date have been limited to binaries containing red giants with periods longer than
1--2 months and shorter than 5-20 years. The lower limit is set by selection effects caused by
possible mass transfer between the binary components (see Pryor, Latham \& Hazen 1988), whereas the 
upper cutoff is simply a consequence of the limited duration (and velocity accuracy) of existing 
radial velocity surveys. 
In this paper, we combine previously published radial velocities for red giants in the nearby
cluster M22 with new measurements accumulated between 1972 and 1994 to search for spectroscopic binaries
with periods in the range 0.2 to 40 years, a factor of two improvement over existing surveys in the upper
period cutoff. In a companion paper (C\^ot\'e \& Fischer 1996),
we report on an investigation of the {\sl short-period} end of the globular cluster binary distribution
(i.e., 2 day $\lae$ P $\lae$ 3 years) based on radial velocities for stars on the upper 
main-sequence of the nearby cluster M4.

\section{Observations}

\subsection{Photometry}

As described in \S 3, our technique of determining $x_b$ from a series of Monte-Carlo simulations 
requires an estimate of the radius of each star in order to assess the likelihood of Roche-lobe overflow
on an object-by-object basis.
Such radii are most easily obtained by comparing the location of each star in the cluster 
color-magnitude diagram (CMD) with the appropriate isochrone (see C\^ot\'e et al. 1994 for details).
However, six of the red giants in our survey have no published magnitudes or colors since they are located 
either at large distance from the cluster center or in the crowded core of the cluster --- 
areas which were specifically avoided in previous photometric studies. Moreover,
the available photometry for the remaining objects (almost all of which is photographic in nature) 
must be assembled from a number of different sources (e.g. Peterson \& Cudworth 1994; 
Lloyd Evans 1975; Arp \& Melbourne 1959).

In order to obtain a homogeneous set of magnitudes and colors for {\sl all} stars
in our survey,
we used the KPNO 0.9m telescope on 1995 July 8/9 with the T2KA 2048$\times$2048 CCD
(scale = 0$\farcs$68 pixel$^{-1}$, gain = 10.7 e$^-$ ADU$^{-1}$, readnoise = 4 e$^-$)
to image five overlapping fields in the direction of M22.
We obtained a BV frame pair centered on M22 and four other fields offset by 14$\farcm$3
in the NW, NE, SW and SE directions, giving a total field of view of 0.52 square degrees.
Seeing during the exposures was typically $\simeq$ 1$\farcs$8.
A finding chart for the six stars without previous photometry is given in Figure 1.
After standard preprocessing with 
IRAF,\altaffilmark{4} \altaffiltext{4}{IRAF is distributed by the National Optical
Astronomy Observatories, which are operated by the Association of Universities for
Research in Astronomy, Inc., under contract to the National Science Foundation.}
a CMD for the entire field was derived using DAOPHOT II (Stetson, Davis \& Crabtree 1990) and the 
related software package ALLFRAME (Stetson 1994).
Unfortunately, observing conditions were nonphotometric, making a direct calibration of the photometry impossible.
We instead used 22 isolated, local photoelectric standards from Alcaino, Liller \& Alvarado (1988) 
to calibrate the data.  Figure 2 shows a comparison between our CCD photometry and the photographic
data of Peterson \& Cudworth (1994) for the 137 radial velocity members which are common to both studies
(see \S 2.2).
There is a tendency for our V magnitudes to be 0.05 mag
fainter than those of Peterson \& Cudworth (1994).
We also find a similar offset in color, ${\Delta}$(B--V) = 0.06 mag,
in the sense that our derived colors are bluer than the photographic values.
The sense of the ${\Delta}$V offset from the Peterson \& Cudworth (1994) photometry 
is the same as that reported in the recent CCD
study of the cluster core by Anthony-Twarog, Twarog \& Craig (1995), although not quite as large: 
0.05 mag compared to 0.09 mag. Unfortunately, since Anthony-Twarog,Twarog \& Craig (1995) did not obtain 
B photometry for their program stars, a direct comparison of the CCD colors is impossible.

The CMD for the entire 43$\farcm$2$\times$43$\farcm$2 field is shown in the upper panel of Figure 3. As expected,
the field star contamination near M22 ($l$ = 9$\fdg$9, $b$ = --7$\fdg$6) is severe.
The lower panel of Figure 3 shows those stars within
five core radii ($r_c$ = 1$\farcm$42 according to Trager, King \& Djorgovski 1995) of the Shawl \& White (1986) cluster center. 
The large open squares indicate the three known photometric
variables in our sample, for which we have adopted the mean magnitudes and colors reported by 
Peterson \& Cudworth (1994). The small squares indicate the remaining 106 
cluster red giants with multiple velocity measurements.

\subsection{Spectroscopy}

An observing log of all spectroscopic observations used in our survey is given
in Table 1, whose columns record, from left to right, the date of the observing run, the
telescope and accompanying instrumentation, the number of separate velocity measurements,
and the total number of stars which were observed.
We now discuss each of these data sets in turn.

As one of the nearest globular clusters (e.g., Djorgovski 1993 quotes a distance
of 3 kpc), M22 was an early target for high resolution spectroscopy of individual cluster giants.
In their landmark dynamical study of M3, Gunn \& Griffin (1979) allude to the existence of 
several dozen, unpublished radial velocities for stars in a number of other clusters, including M22,
which were collected with the Hale 5m telescope and Palomar radial velocity scanner during the 1970s. 
The entire sample of Hale radial velocities for M22 giants (67 observations of 44 different stars), 
accumulated during a series of observing runs in 1972, 1974 and 1975, were kindly donated by Drs. 
Griffin and Gunn.
For a complete description of the design and operation of the Palomar radial velocity scanner,
the reader is referred to Griffin \& Gunn (1974) and Gunn \& Griffin (1979).

In addition to these data, a total of 176 radial velocities for 130 red giants in M22 have recently been 
published by Peterson \& Cudworth (1994). These velocities, accumulated during the 1985 -- 1987
observing seasons with the SAO 1.5m and MMT telescopes, have a precision similar to the 
Hale observations (e.g., typical uncertainty $\simeq$ 1 km s$^{-1}$). For a detailed
description of the SAO observations and reductions, the reader is referred to
Peterson \& Cudworth (1994).
We obtained a third set of velocities using the
DAO radial velocity scanner (Fletcher et al. 1982) on the CFHT during a pair of observing runs in 1989 and 1991.
A total of 49 radial velocities for 36 stars (chosen from the finder charts of Lloyd Evans 1975 and Cudworth 1986) 
were measured over three nights.
The velocity accuracy delivered by this combination of telescope and instrument 
is also $\simeq$ 1 km s$^{-1}$. A complete description of this instrument's operation 
at the coud\'e focus of the CFHT may be found in Pryor et al. (1989).
Our final epoch consists of radial velocities (median precision $\simeq$ 1.6 km s$^{-1}$) for 
92 giants obtained with the CTIO 4m telescope and the Argus multi-object spectrograph in June 1994.
Targets were selected from the list of confirmed cluster members given in Peterson \& Cudworth (1994).
The primary goal of this observing run was to measure velocities for turnoff dwarfs
in the nearby globular cluster M4, the results of which are presented by
C\^ot\'e \& Fischer (1996). Since the reduction procedures followed in the M22 analysis
are identical to those used in our study of M4, the reader is referred to that paper 
for details.
 
To ensure that all of these radial velocities share a common zeropoint, we have applied small offsets
to the Hale, CFHT and CTIO observations. These corrections, which
never exceeded 1 km s$^{-1}$, were determined by matching all stars in common with the 
SAO/MMT catalog (which is the most extensive data set and has the most closely monitored zeropoint).
The entire sample of M22 radial velocities is given in Table 2 
(also presented in the ApJ/AJ CD-ROM Series, Volume X, 1996).
Since these data may be useful for other purposes, we have tabulated {\sl all} known radial velocity
members of M22; that is to say, cluster members with only a single radial velocity measurement are
included.  Table 2 therefore consists of 383 radial velocities for 162 different stars. 
Of these, 109 objects have more than one measurement. 
Excluding the three known photometric variables in this sample (i.e., stars V5, V8, and V9; Sawyer-Hogg 1973),
we find a systemic velocity of -148.55$\pm$0.56 km s$^{-1}$ and a one dimensional velocity dispersion 
of 7.06$\pm$0.40 km s$^{-1}$ using the maximum-likelihood method of Pryor \& Meylan (1993).
From left to right, Table 2 gives the star number, the ID from Peterson \& Cudworth (1994),
the distance in arcminutes from the Shawl \& White (1986) cluster center, the position angle in degrees, 
the heliocentric Julian date, the radial velocity, the weighted mean velocity, 
the reduced chi-squared and the probability P$(\chi^2)$ of that chi-squared value being exceeded assuming that the
velocity is constant, the V magnitude and B--V color measured 
from our CCD frames, and the source of the radial velocity (P = Palomar, S = SAO, C = CFHT, and T = CTIO). 

Most previous surveys of red giants in globular clusters have found evidence
for a velocity ``jitter" which is assumed to arise from convective or pulsational motions in the
atmospheres of these evolved stars (Gunn \& Griffin 1979; Mayor et al. 1984; Lupton, Gunn \& Griffin 1987; Pryor, 
Latham \& Hazen 1988). 
Gunn \& Griffin (1979) found it necessary to included an external dispersion of
0.8 km s$^{-1}$ to account for the low-amplitude velocity variations observed in their sample of M3 giants.
Using an expanded sample of M3 velocities, Pryor, Latham \& Hazen (1988) concluded that, while the
velocity variability was probably present in the entire sample of program objects, it was certainly
largest for those stars within 0.5 magnitude of the giant-branch tip.
We too find evidence for such a luminosity-dependent velocity jitter in our sample of M22 giants. 
Figure 4 shows the dependence of P$(\chi^2)$ on V magnitude and B--V color for the 106
stars with multiple velocity measurements (photometric variables excluded).
No jitter has been assumed in deriving the probabilities shown in Figure 4.
Four stars with P$(\chi^2) \le 0.001$ are indicated by the vertical arrows.
As Figure 4 demonstrates, there is a tendency for stars near the tip of the giant branch
to have the lowest probabilities.
This effect can also be seen by dividing our sample into three broad luminosity bins: a bright sample
of 13 stars (V $\le$ 11.5), an intermediate sample of 32 stars (11.5 $\le$ V $\le$ 12.4) and
a faint sample (12.4 $\le$ V $\le$ 13.65) of 61 stars. The respective number of objects with
P$(\chi^2) \le 0.5$ in each of these three bins are 11 (85\%), 22 (69\%), and 30 (49\%), which demonstrates that the size of
the jitter is indeed luminosity-dependent.

After some experimentation, we adopted a velocity jitter which is constant for those stars 
within $\simeq$ 0.5 magnitude of the tip of the RGB and which linearly decreases to zero
below this level:
$${\sigma}_j = 0.9~{\rm km~s^{-1}~~~~~~~~~~~~~~~~~~~~~~~~~~~~~~~~~~~~~~~~~~~~~~~~~~~~~~~~~~~~~if~V \le 11.55}$$
$${\sigma}_j = (0.9/2.45)(14.0-{\rm V})~{\rm km~s^{-1}~~~~~~~~~~~~~~~~~~~~~~~~~~~~if~11.55 < V \le 14.00}\eqno{(3)}$$
$${\sigma}_j = 0~{\rm km~s^{-1}~~~~~~~~~~~~~~~~~~~~~~~~~~~~~~~~~~~~~~~~~~~~~~~~~~~~~~~~~~~~~~~~if~V > 14.00}$$
Adding this jitter in quadrature to the formal velocity uncertainties given in Table 2
lowers the number of stars with P$(\chi^2) \le 0.5$ in each of the three luminosity bins to
8 (62\%), 13 (41\%), and 30 (49\%), respectively. The resulting histogram of 
P$(\chi^2)$ is given in the upper panel of Figure 5.
For a sample of constant-velocity stars, such a distribution should be nearly
flat (provided, of course, that the velocity uncertainties are correctly estimated). A population of
stars with variable velocities will manifest itself as a peak near zero probability.
Apart from a modest enhancement near P$(\chi^2) \approx 0.00$, the histogram is clearly
consistent with a flat distribution (indicated by the dashed line).

The heavy line in the lower panel of Figure 5 shows the cumulative distribution of 
P$(\chi^2)$. The distribution expected for a sample of constant-velocity
stars is indicated by the dashed diagonal line. This technique avoids the problem of binning
the data which is inherent in the above approach (although it should be kept in mind that
the individual points are no longer statistically independent). 
The good agreement between the two distributions lends credence to
our adopted jitter and suggests that it is unlikely that our M22 sample contains an 
appreciable number of bona fide radial velocity variables.
As a further demonstration of the validity of our adopted velocity jitter, we show the probability
distributions for the two cases of: (1) no velocity jitter (upper line); and (2) 
a jitter of 1.5 km s$^{-1}$ which is independent of luminosity (lower line). 
These distributions bracket that obtained using equation 3 and
provide a much poorer match to the distribution expected for
a sample composed primarily of constant-velocity stars.
Regardless of the precise size and luminosity dependence of the external dispersion, 
our estimate for the cluster binary fraction is based solely on the number of stars showing
velocity variations greater than 8 km s$^{-1}$. As a result, our best-fit binary fraction is
essentially independent of the adopted jitter (see \S 4).

\section{Monte-Carlo Simulations}

Extracting a binary fraction from a catalog of radial velocities is a formidable task,
since the number of stars expected to show large radial velocity variations depends on
the number, spacing, and precision of the observed radial velocities,
as well as the distribution of orbital periods, mass ratios, eccentricities, inclinations,
and the longitudes and times of periastron passage. Since 
an {\sl a priori} knowledge of the form of these distributions is, of course, unavailable,
the best approach is to generate simulated radial velocity catalogs
with known binary fractions. By comparing the simulated observations to the actual data,
it is possible to determine both $x_b$ and its associated confidence range.
The Monte-Carlo approach used here is similar to that employed in several earlier 
studies (Hut et al. 1992; C\^ot\'e et al. 1994), to which the reader
reader is referred for more details.

First, we compute a radius for each star based on its location in the CMD
using the 14 Gyr, [Fe/H] = --1.78 isochrone of Bergbusch \& VandenBerg (1992). 
(We adopt a distance of 3 kpc and a metallicity of [Fe/H]  = --1.75; Djorgovski 1993.) 
Figure 6 shows the Bergbusch \& VandenBerg (1992) radius-magnitude relation used 
to compute the radii R of our stars.
Second, for each simulation, we {\sl assume} a binary fraction and randomly assign binary or single
star status to each program object using $x_b$ as the probability of selecting a binary.
For single stars, we generate
the same number of radial velocities as in the actual catalog and include a realistic amount of 
observational noise (i.e., ${\sigma}^2$ = ${\sigma_{\rm f}}^2$ + ${\sigma_{j}}^2$ where 
${\sigma_{\rm f}}$ is the formal uncertainty recorded in Table 2 and ${\sigma_{j}}$ is the
external dispersion given by equation 3).
For binary stars, we randomly assign an orbital period P and mass ratio, q = M$_2$/M$_1$. 
We adopt distributions for P and q which are based on those found by Duquennoy \& Mayor (1991)
in their study of multiplicity among nearby, solar-type stars.
Specifically, we adopt a period distribution of the form 
$${d{\rm N} \over d{\rm \log{P}}} \propto \exp(-{(\log{\rm P} - \overline{\log{\rm P}})^2 \over 
{2\sigma_{\rm log~P}^2}}),\eqno{(4)}$$
where P is in days, $\overline{\log{\rm P}}$ = 4.8 and $\sigma_{\rm \log{P}}$ = 2.3.
For the distribution of mass ratios, we adopt the Duquennoy \& Mayor (1991) relation
$${d{\rm N} \over d {\rm q}} \propto \exp(-{({\rm q} - \overline{\rm q})^2 \over {2\sigma_{\rm q}^2}}),\eqno{(5)}$$
where $\overline{\rm q} =$ 0.23 and $\sigma_{\rm q}$ = 0.42.
Our adopted upper and lower cutoffs for these distributions are discussed below.

Simulations are carried out for two
assumed eccentricity distributions: (1)  purely circular orbits, $e = 0$; and
(2) a thermal distribution of eccentricities, $f(e) = 2e$ (Heggie 1975).
The final step is to assign random values to each of the remaining orbital parameters
and test for the possibility of mass transfer between
the components using the prescription given in Pryor, Latham \& Hazen (1988). If the binary
is found to be sufficiently compact that Roche-Lobe overflow is likely, we assume that the
system has been removed from the sample (see Pryor, Latham \& Hazen 1988) and repeat the entire procedure.
If mass transfer has {\sl not} occurred, 
we generate the appropriate number of radial velocities 
for each star, including, as before, a realistic amount of observational noise.
For a grid of $x_b$ running from 0.00, 0.01, 0.02,..., 0.60 we generate 1000 simulated catalogs 
and calculate the mode of the number of stars which show velocity variations greater than 8 km s$^{-1}$,
a value chosen to avoid possible complications caused by measurement error and/or atmospheric motions
(which can sometimes approach 8 km s$^{-1}$ for long-period variables near the tip of the giant branch; 
Hut et al. 1992). 

Needless to say, a crucial first step in the analysis is to determine the range of orbital periods
and mass ratios to which our observations are sensitive. We measure our {\sl binary 
discovery efficiency} by determining the fraction of known binary stars in each simulation
which are recovered as binaries by exhibiting velocity variations greater than 8 km s$^{-1}$. 
Figure 7 shows our binary discovery efficiencies as a function of orbital period for the two cases of circular 
and thermal orbits.
To illustrate the effects of mass transfer between the components, we
display the discovery efficiencies before (solid curves) and after (dashed curves) taking mass
transfer into account. We conclude that our survey should be sensitive to binaries with
$0.1 \le$ q $\le 1.0$ and 0.2 $\lae$ P $\le$ 40 years,
where the approximate lower limit on P reflects the fact that the actual cutoff used in the simulations was 
0.1 years for circular orbits and 0.3 years for thermal orbits.

\section{Results}

Only one of our program stars (V-23) shows a velocity variation larger than 8 km s$^{-1}$. 
The reality of this variation, however, is questionable since
four of the five velocities for this star (which span nearly 20 years) are in good
agreement at $v_r$ $\simeq$ --138.5 km s$^{-1}$. The remaining velocity --- that obtained in July 1989 with the 
CFHT --- is in disagreement by more than 12 km s$^{-1}$ 
(i.e., $v_r$ = --150.76 km s$^{-1}$). Moreover, V-23 is located in the crowded 
core of the cluster and just 10$^{\prime\prime}$ from the star I-202 which has 
$v_r$ = --151.85 km s$^{-1}$ based on a single measurement with the Hale 5m telescope.
We therefore suspect that V-23 may have been misidentified during the July
1989 CFHT run, although continued monitoring of this star is clearly in order since,
at present, we cannot rule out the possibility that it is an eccentric binary.
In what follows, we give the binary fractions which result if we: (1) {\sl discard}
the July 1989 measurement (Case A); and (2) {\sl retain} the July 1989 measurement (Case B). 
For Case A, the observed velocity variation of V-23 drops to 2.00 km s$^{-1}$. The stars
with the largest velocity variations in the Case A sample are then III-35 and IV-97 which shows changes
in radial velocity of 6.92 km s$^{-1}$ and 6.90 km s$^{-1}$, respectively.

For each simulated dataset, we count the number of stars, N$_8$, which show a
velocity variation greater than 8 km s$^{-1}$. The procedure is repeated 1000
times for each $x_b$ and we take the mode of the resulting distribution as the
value of N$_8$ appropriate for the adopted binary fraction.
We then determine the cluster binary fraction by finding the $x_b$ for which
the {\sl observed} and {\sl simulated} values of N$_8$ are equal.
To get the 90\% confidence limits on the derived binary fraction, we find: (1) the {\sl smallest} value of 
$x_b$ which produces a value of N$_8$ which equals or exceeds the actual value less than 5\% of the time and;
(2) the {\sl largest} value of $x_b$ which gives
a value of N$_8$ which is equal to or less than the actual value less than 5\% of the time.
For Case A, we find acceptable matches (i.e., N$_8$ = 0) between the simulated and actual data 
for binary fractions in the range $0.00 \le x_b \le 0.01$ (circular orbits) and $0.00 \le x_b \le 0.03$ 
(thermal orbits). The 90\% confidence intervals for these estimates
are $0.00 \le x_b \le 0.11$ and $0.00 \le x_b \le 0.19$, respectively.
The binary fractions for Case B (i.e., N$_8$ = 1) are $0.02 \le x_b \le 0.05$ (circular orbits) and 
$0.04 \le x_b \le 0.12$ (thermal orbits), with respective 90\% confidence intervals of
$0.00 \le x_b \le 0.18$ and $0.00 \le x_b \le 0.29$. These values of $x_b$ refer to the fraction of primaries on the
main sequence with 0.2 $\lae$ P $\le$ 40 years and 0.1 $\le$ q $\le$ 1.0.

How sensitive are these estimates to our adopted model parameters? To answer
this question we have derived $x_b$ and its corresponding 90\% confidence limits 
using a number of different model assumptions.
The results of these experiments are summarized in Table 3.
Specifically, we have investigated the effect of: 
(1) increasing the lower cutoff of the secondary mass distribution (model b);
(2) increasing the upper cutoff in the secondary mass distribution (equivalent to including
some massive degenerate secondaries) (model c);
(3) selecting periods and mass ratios from logarithmically {\sl flat} distributions (model d);
(4) disregarding the presence of velocity jitter (model e);
(5) decreasing the upper period cutoff by a factor of two (model f); and
(6) selecting only equal-mass components (model g).
In general, the changes in the
derived binary fractions produced by these various assumptions are small compared
to the 90\% confidence intervals on $x_b$. The sole exception is model g, which assumes
equal mass components. In this case, the upper limits on the derived binary fraction
drop by nearly a factor of two since binaries with equal mass components show relatively large
velocity variations and are, therefore, easier to detect.
In short, the two factors which limit the precision of existing measurements of $x_b$ 
based on the velocities of red giants remain
the unknown distribution of orbital eccentricities and
the statistical uncertainties due to small samples of stars.
Considerable progress toward overcoming the second of these difficulties is
expected in the next few years, as repeat velocities for hundreds, or in some cases, 
thousands of globular cluster stars are accumulated with multi-object 
spectrographs (e.g., Gebhardt et al. 1995).

We now compare our best-fit (i.e., Case A) binary fractions of $x_b$  = 0.01$^{+0.10}_{-0.01}$ (circular orbits)
and $x_b$ = 0.03$^{+0.16}_{-0.03}$ (thermal orbits)
to that found among nearby field stars. The most comprehensive study of multiplicity among
Population I field stars is that of Duquennoy \& Mayor (1991),
who surveyed 164 primaries in the spectral range F7 to G9. Their complete sample contains 21 binaries with periods
between 0.1 and 40 years (our detection limits for circular orbits) and 20 systems with periods
in the range 0.3 to 40 years (our limits for thermal orbits). If we add one star to account for
incompleteness and multiply by 0.88 to remove binaries having mass ratios less than 0.1,
we find $x_b \simeq$ 0.12$\pm$0.03. This is the Population I binary fraction to be compared to that of M22.
This is considerably larger than that found for M22, though we caution that
the 90\% confidence intervals on the derived binary fraction still overlap the Population I estimate.

Binary fractions based on radial velocity surveys have now been published for several
other globular clusters, the results of which are summarized in Table 4a.
The first four columns of this table record the name of the cluster(s),
the binary fraction for systems with mass ratios in the range 0.1 $\le$ q $\le$ 1.0, 
the shortest and longest periods detectable in the survey (P$_{\rm min}$ and P$_{\rm max}$, respectively), 
and $x_b$/log(P$_{\rm max}$/P$_{\rm min}$), the measured binary fraction per decade of period
(which has the advantage that it removes, to some extent, the dependence  of $x_b$
on the adopted period range).
We begin the discussion of these cluster binary fractions by comparing
them to the value reported by Yan \& Mateo (1994) based on the five
eclipsing binaries that they found in M71: $x_b=0.013$ for systems
with periods in the range 2.5 to 5 days.
Using their binary fraction to extrapolate to the period interval sampled by the radial velocity surveys 
reviewed by Hut et al. (1992), these authors concluded
that ``existing data on short-period and red-giant binaries favor a flat frequency-period distribution over
the distribution given in Duquennoy \& Mayor (1991)."
However, with a one-dimensional velocity
dispersion of $\sigma_{\rm 1D}$ $\simeq$ 2.2 km s$^{-1}$, M71 represents a dynamical environment which is quite different from 
the bulk of the clusters in the Hut et al. (1992) sample (which have $\overline{\sigma}_{\rm 1D}$ $\simeq$ 6.3 km s$^{-1}$; see below).
Ideally, one would like to compare the abundances
of short- and long-period binaries in the {\sl same} cluster or, at least, in
clusters having similar velocity dispersions and stellar densities. 
It is also worth bearing in mind that both the Yan \& Mateo (1994) and Hut et al. (1992) binary
fractions are uncertain by roughly a factor of two: the former as a result of the poorly known coalescence
timescales of contact binaries and the latter as a result of the low binary discovery efficiencies for luminous giants.

To investigate whether or not the conclusions of Yan \& Mateo (1994) are supported by the expanded sample 
of clusters in Table 4a,
we have ``predicted" binary fractions for the various
radial velocity surveys using the Yan \& Mateo (1994) short-period binary fraction and 
assuming: (1) a period distribution which is flat in $\log{\rm P}$; and
(2) a period distribution identical to that observed by Duquennoy \& Mayor (1991) for nearby solar-type
stars (i.e., equation 4). The respective estimates, $x_b^\prime$(flat) and $x_b^\prime$(DM91),
are recorded in the fifth and sixth columns of Table 4a.
For M71, the measured binary fraction agrees within its uncertainty with both the flat and DM91 distributions. In other
words, M71 appears to be relatively abundant in {\sl both} short- and long-period binaries. Similarly, both the 
flat and DM91 extrapolations are consistent with the measured binary fraction in M4 (which is a similar dynamical
environment to M71). 
On the other hand, both extrapolations overpredict $x_b$ for M22 and $\omega$ Cen,
clusters which have surveys sensitive to longer periods and considerably higher binary ionization rates (see below).
Indeed, for these clusters, $x_b$ and $x_b^\prime$(DM91) differ by nearly an order of magnitude.
We conclude that, based on the expanded sample of clusters in Table 4a, the overabundance of 
short-period, eclipsing binaries compared to red-giant binaries
noted by Yan \& Mateo (1994) may, in fact, be stronger evidence for the disruption of dynamically-soft 
binaries rather than for a universal period distribution 
which is flat in $\log{\rm P}$.

Is there evidence for the disruption of soft binaries in the radial velocity surveys alone? 
Table 4b records the clusters which have been searched for spectroscopic binaries, the binary fraction 
per decade of period, the logarithmic mean period of the survey,
the one-dimensional velocity dispersion calculated with the maximum-likelihood estimator of Pryor \& Meylan (1993) 
{\sl and measured at the same radius as those stars monitored for velocity variability,}\altaffilmark{5}\altaffiltext{5}{For 
the clusters reviewed in Hut et al. (1992), we give
the mean $\sigma_{\rm 1D}$, weighted by the number of stars in each cluster.} the 
critical binary separation a$_{\rm c}$ estimated from equation 1, and the corresponding critical 
binary period P$_{\rm c}$ for a pair of 0.8M$_{\odot}$ stars.
In Figure 8 we show the binary fraction per decade of period plotted against: (1) the logarithmic mean period of
each survey; (2) the critical binary separation; (3) the critical binary period; and (4) the ratio of 
the critical binary period to the logarithmic mean period of the survey. 
Although the uncertainties in the measured binary fractions
remain rather large, these figures
demonstrate that M22 and $\omega$ Cen --- clusters which have the largest P$_{\rm max}$ and the smallest P$_{\rm c}$ ---
have the lowest binary fractions.
Similarly, those clusters with the largest
P$_{\rm c}$/$<$P$>$ (e.g., M71, M4 and NGC 3201) have the highest binary fractions in the sample.
It is, of course, possible that {\sl external} processes have modified the binary fraction in a few
of these clusters (for instance, the large number of binaries in M71 might be due, in part, 
to the enhancement of the binary fraction by tidal stripping; McMillan \& Hut 1994).
It should also be kept in mind that the above estimates of a$_{\rm c}$ and P$_{\rm c}$ do not
include {\sl all} of the environmental factors governing binary destruction:
for example, the center of $\omega$ Cen does not appear to satistfy equation 2, nor
does this equation take into account the shrinking of hard binaries with wide orbits 
via energy exchanges with passing stars (see equation 2 of Phinney 1996).
Nevertheless, these values should be a reasonable first approximation to the amount of 
binary destruction expected.  We therefore conclude that both the
number of short-period eclipsing binaries relative to long-period radial velocity variables {\sl and}
the radial velocity surveys alone point to
a frequency-period distribution in which binaries with periods in excess of the ``hard-soft transition" have been
disrupted by stellar encounters (Heggie 1975; Hills 1984).
{\sl To the best of our knowledge, this is the first observational evidence for the 
destruction of soft binaries in globular clusters. }
 
Finally, we note that three stars which have previously been identified as members of the rare breed of 
globular cluster CH stars are included in our
sample: III-106 (Hesser, Hartwick \& McClure 1977; McClure \& Norris 1977), 
IV-24 (Hesser \& Harris 1979) and III-78 
(Lloyd Evans 1978). Continued monitoring of the radial velocities of field CH stars has 
conclusively demonstrated (McClure \& Woodsworth 1990) that many, perhaps even all, of these systems are binaries.
Although the number of velocities for each CH star in our survey is small (III-106 -- three measurements; 
IV-24 and III-78 -- two measurements each), the combined chi-squared for these three
objects is 5.04 for four degrees of freedom, which corresponds to P($\chi^2$) = 0.28. 
Therefore, there is, at present, no evidence that
any of the CH stars in M22 are members of binary systems. A similar result has recently been reported
by Mayor et al. (1996), who 
found only two binaries among 32 chemically peculiar (Ba, CH and S) stars in $\omega$ Cen.
Such a result can be understood if these stars
are simply otherwise normal outliers in the cluster metallicity distribution function
or if they were enriched via mass exchange with companions which have
subsequently been disrupted. However, the field CH stars studied by 
McClure \& Woodsworth (1990) have periods near 3 years, whereas
such systems would be dynamically hard in M22. According to Hills (1984),
binaries with periods less than $\simeq$ 25 years are expected to have escaped disruption, 
for an assumed one-dimensional velocity dispersion of 6.6 km s$^{-1}$ (Peterson \& Cudworth 1994).
Available evidence therefore suggests that the carbon enhancement seen in the M22 CH stars 
is probably {\sl not} the result of mass transfer.
Indeed, Vanture \& Wallerstein (1992) 
have recently carried out an abundance analysis of III-106 and have argued
that it is not a genuine CH star in the sense applied to 
field CH stars. Rather, it appears to be an otherwise normal giant whose 
carbon excess has arisen from incomplete CN processing relative to other M22 giants.

\acknowledgments

We thank Roger Griffin and Jim Gunn for providing the crucial first epoch of radial velocities. 
We also extend our thanks to the staff and night assistants of all three observatories for their 
fine support.  We are especially grateful to Tom Ingerson and Nick Suntzeff for donating a portion 
of an Argus engineering night which was used to obtain some of the radial velocities presented in this 
paper. The research of CP on binary stars in globular clusters was supported by NSF Grant AST-9020685.

\vfill\eject
 
\centerline{~}
\includegraphics{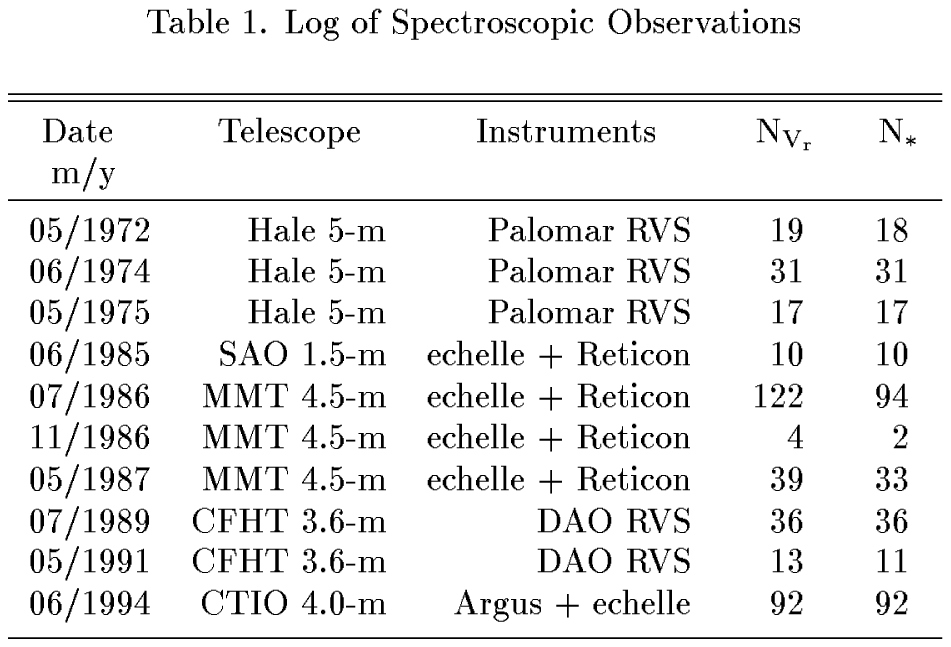}
\vfill\eject
 
\centerline{~}
\includegraphics{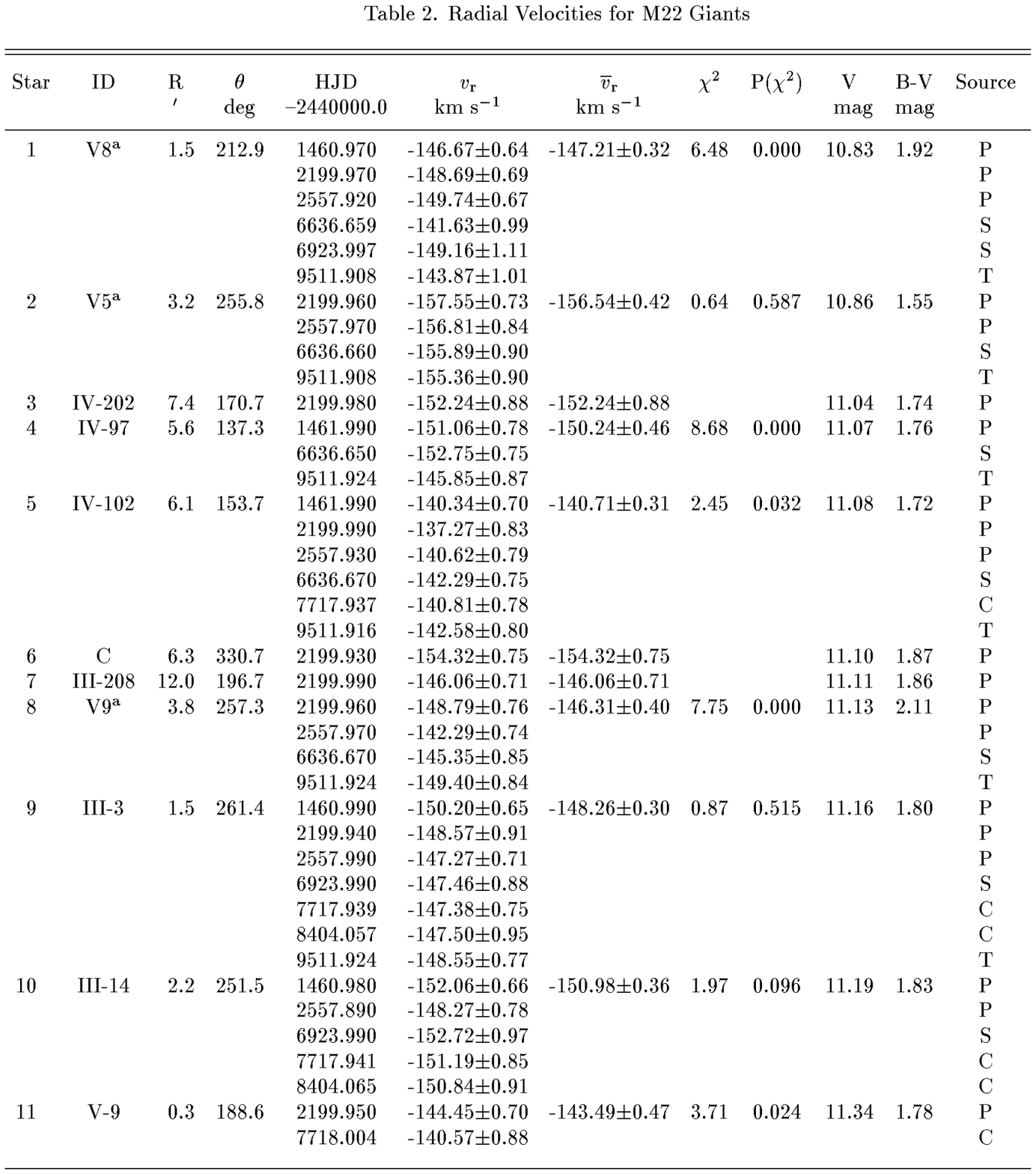}
\vfill\eject
 
\centerline{~}
\includegraphics{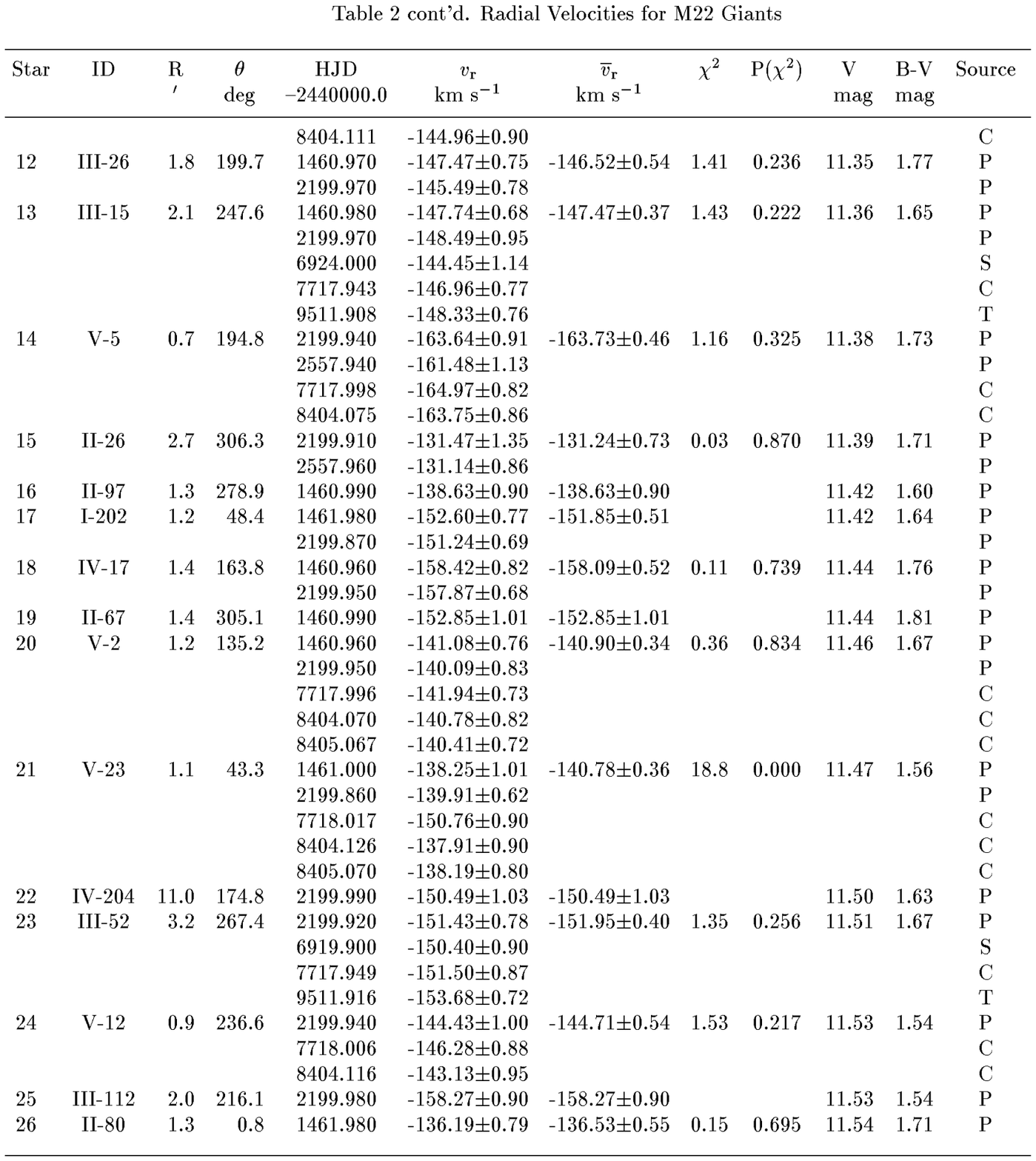}
\vfill\eject

\centerline{~}
\includegraphics{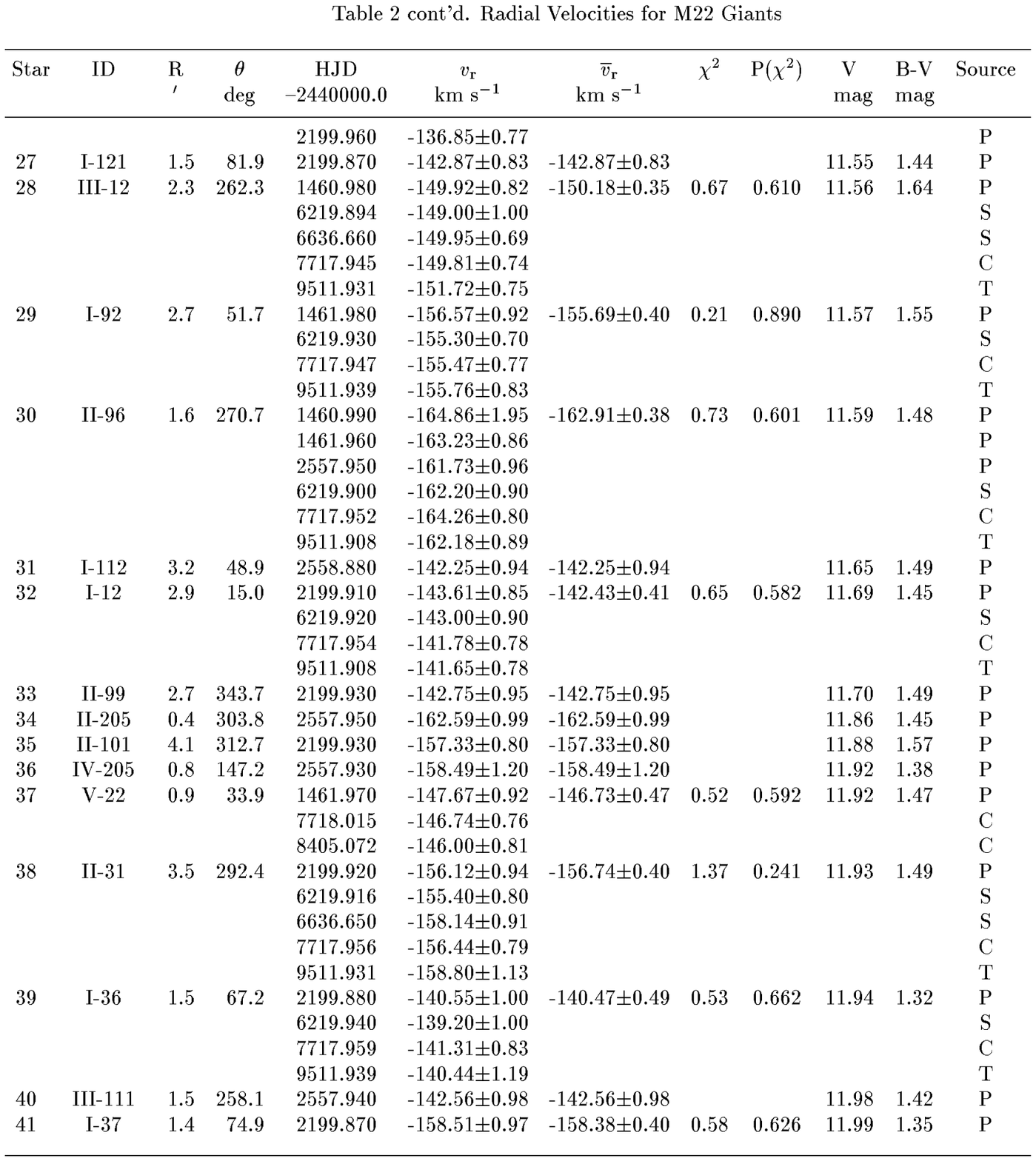}
\vfill\eject
 
\centerline{~}
\includegraphics{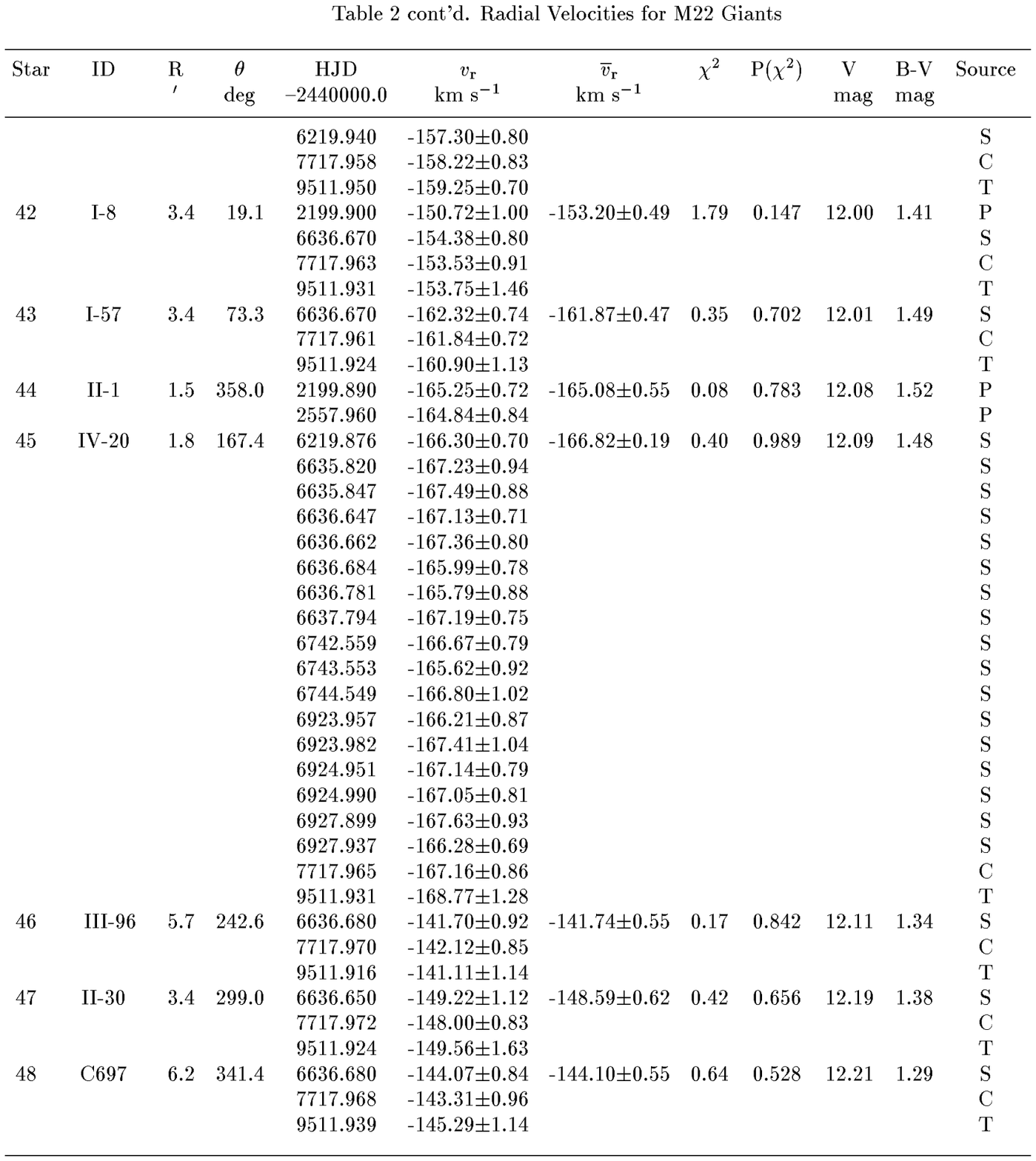}
\vfill\eject

\centerline{~}
\includegraphics{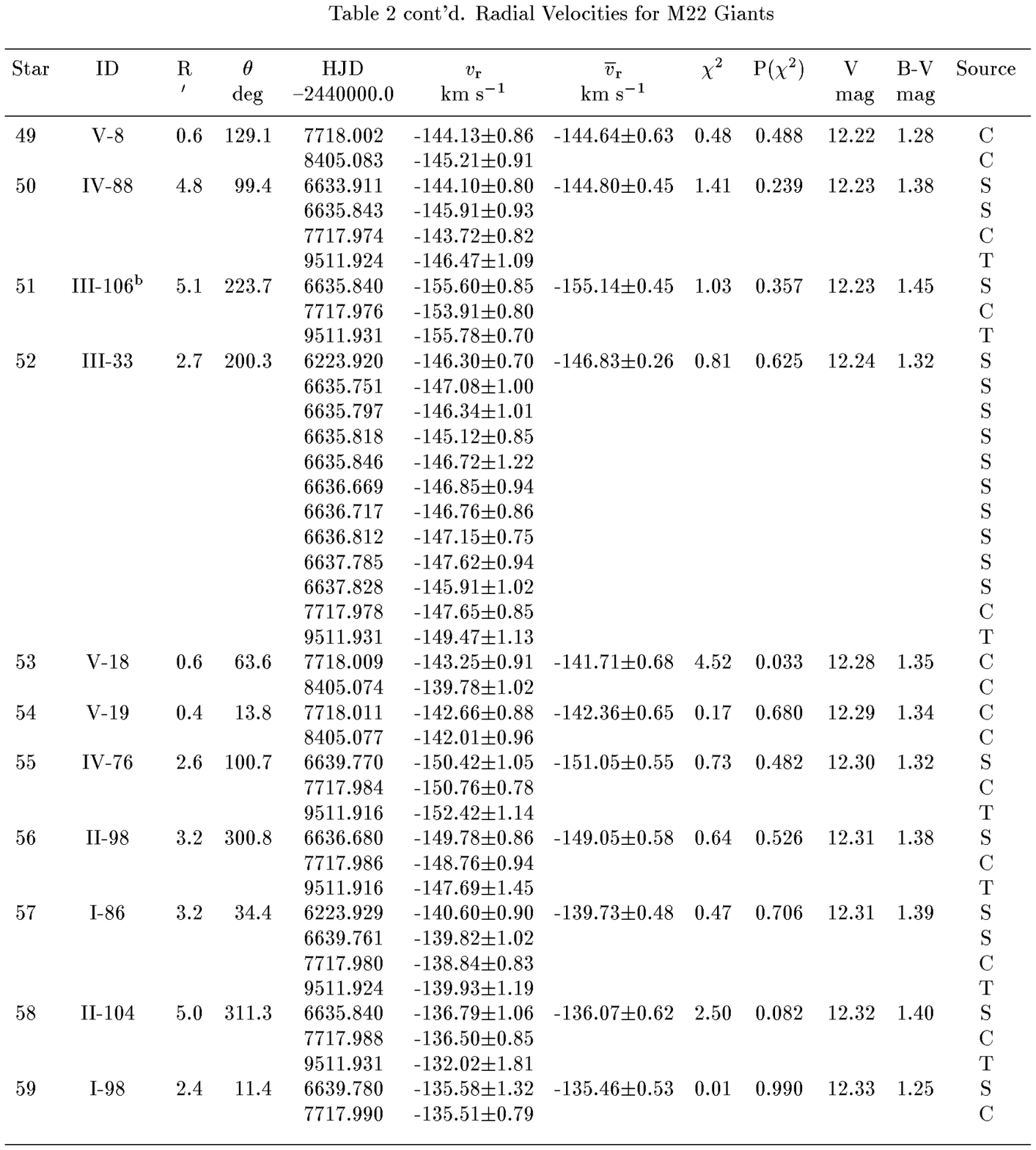}
\vfill\eject
 
\centerline{~}
\includegraphics{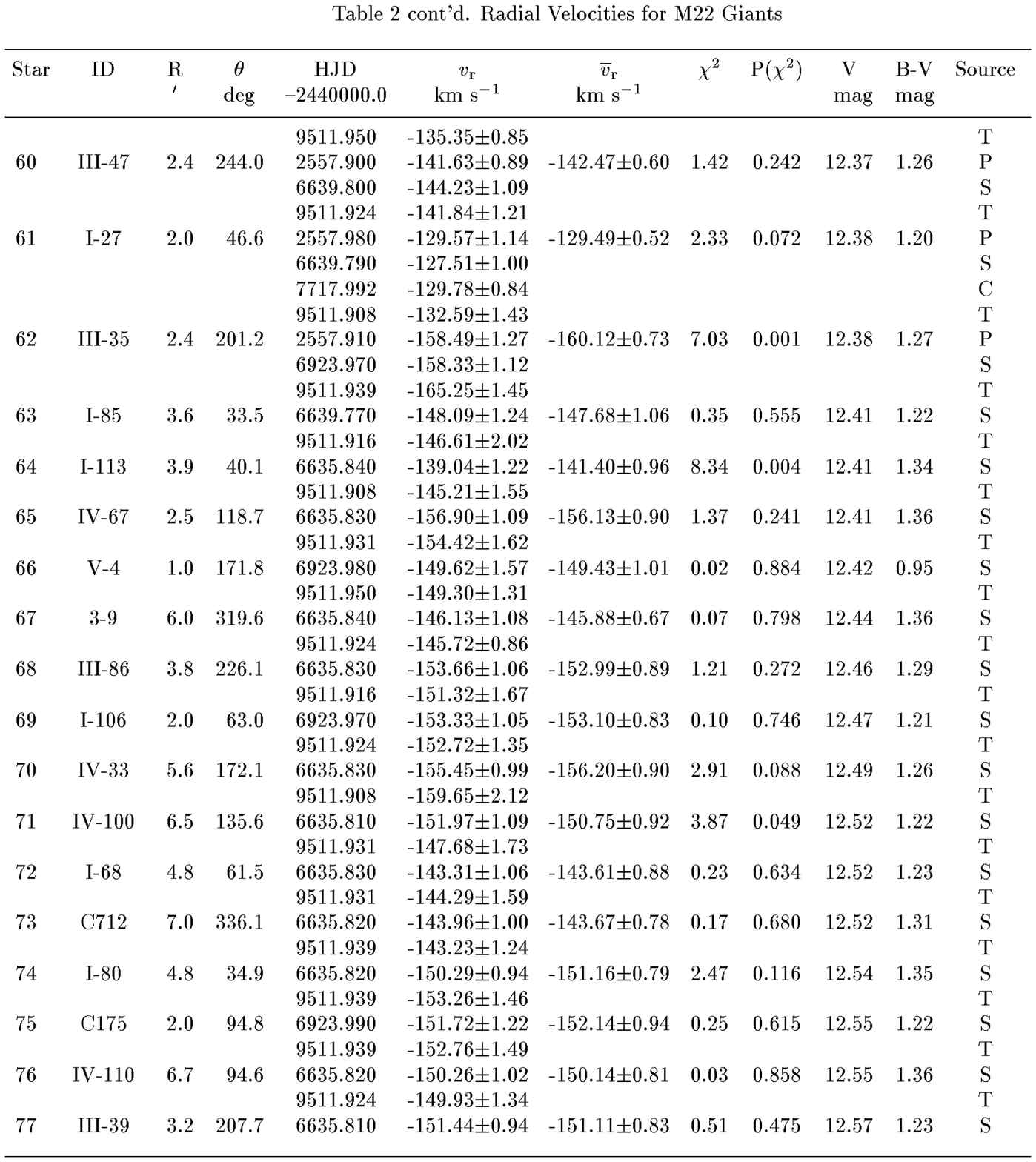}
\vfill\eject

\centerline{~}
\includegraphics{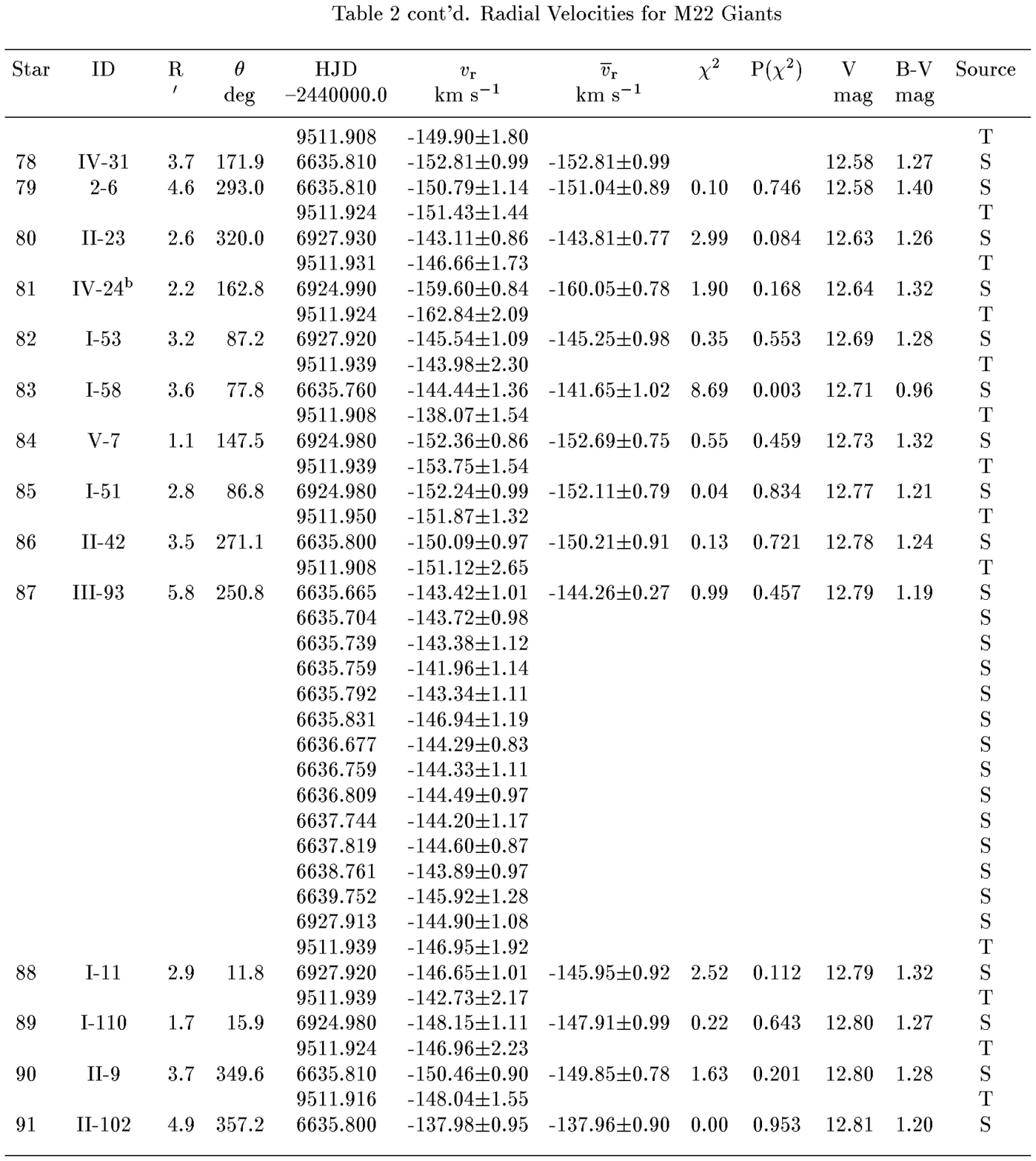}
\vfill\eject
 
\centerline{~}
\includegraphics{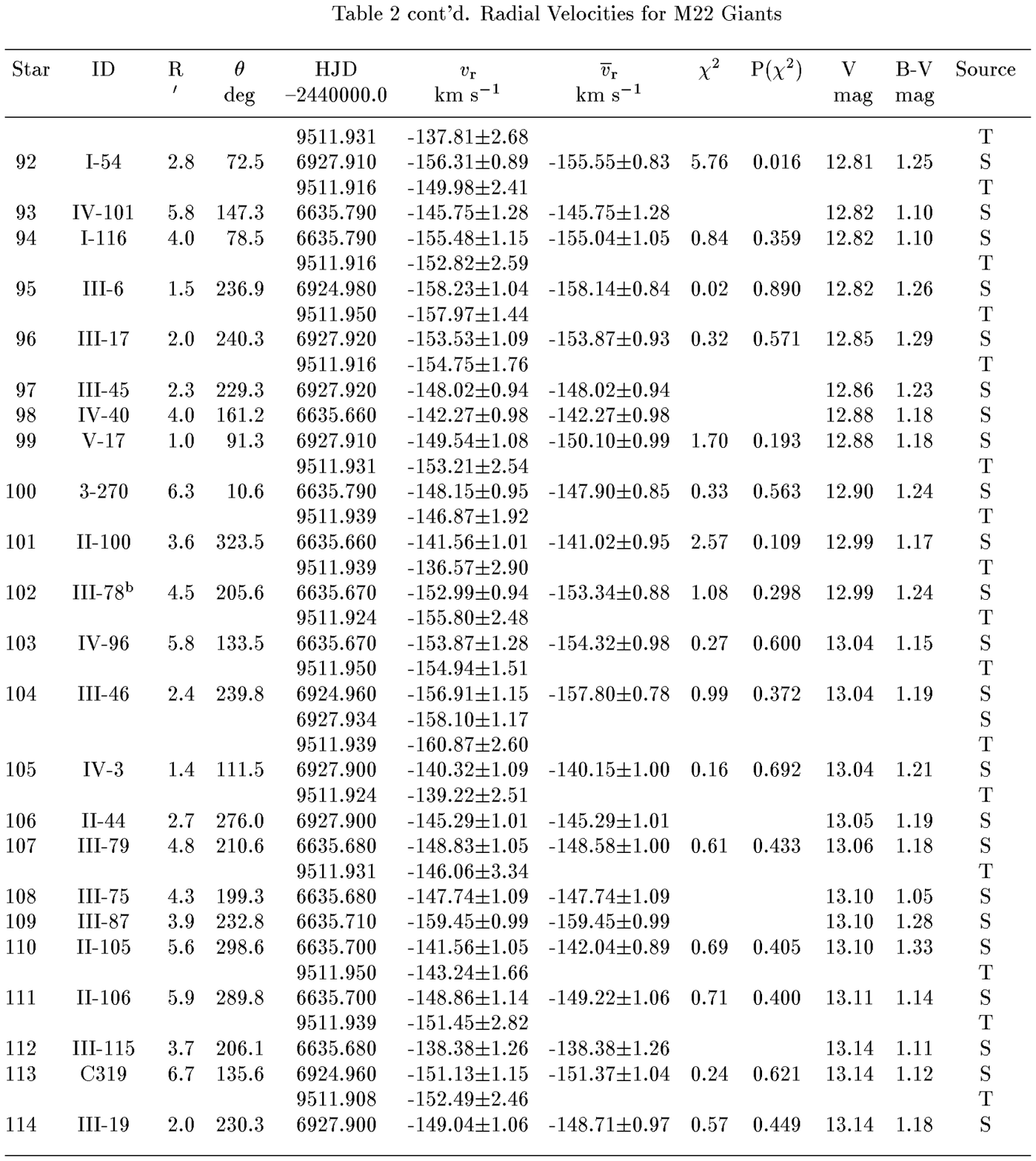}
\vfill\eject

\centerline{~}
\includegraphics{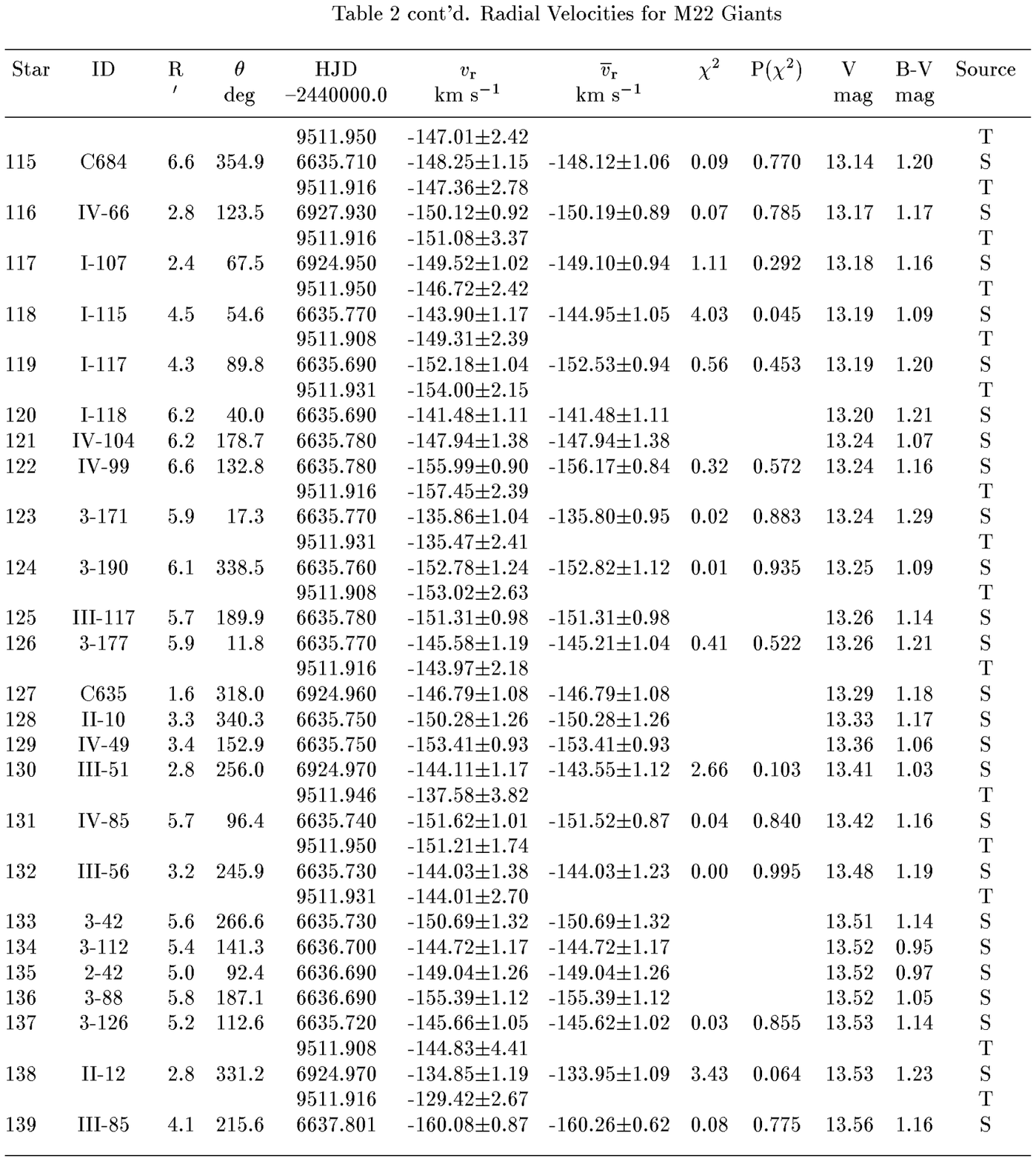}
\vfill\eject
 
\centerline{~}
\includegraphics{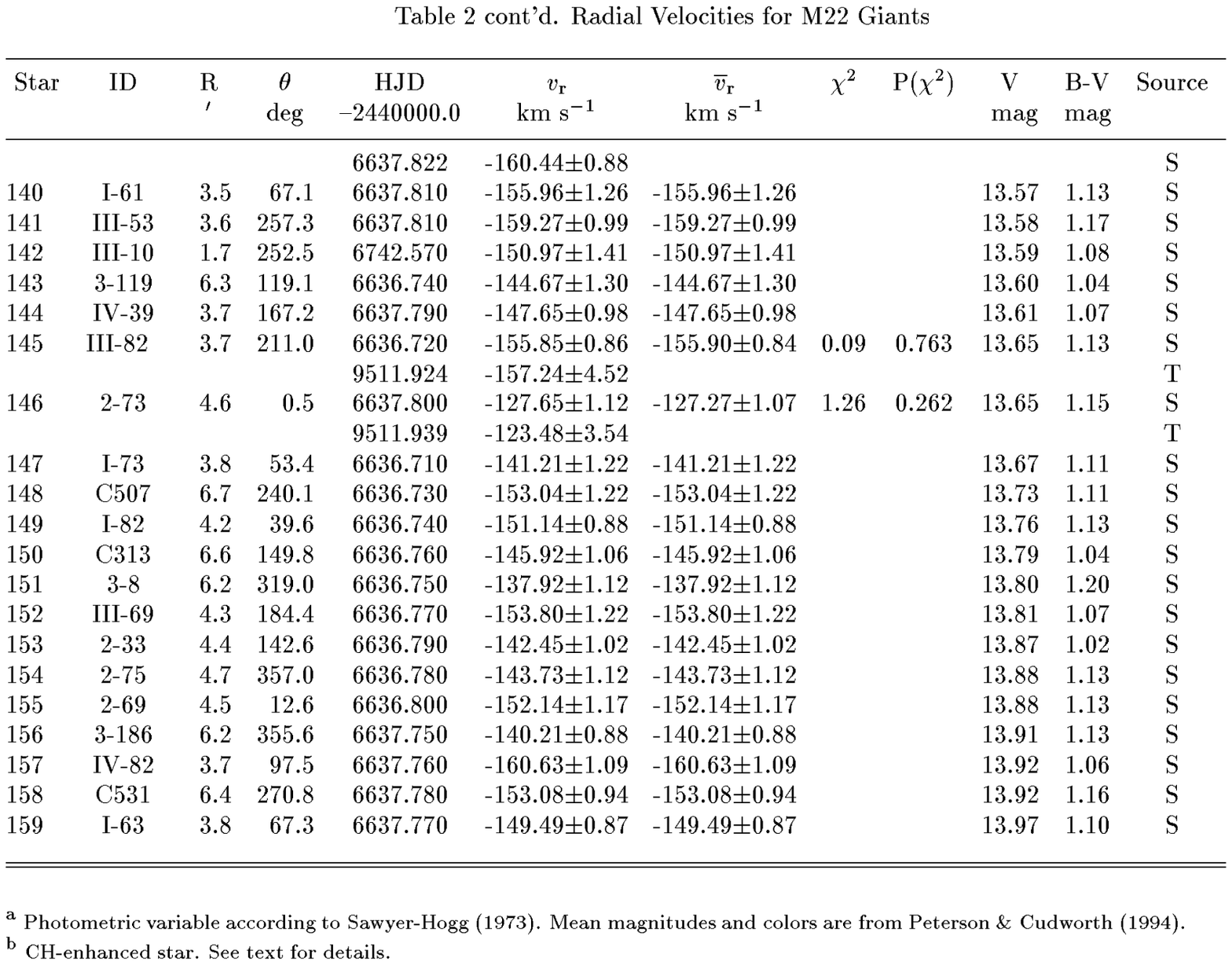}
\vfill\eject

\centerline{~}
\includegraphics{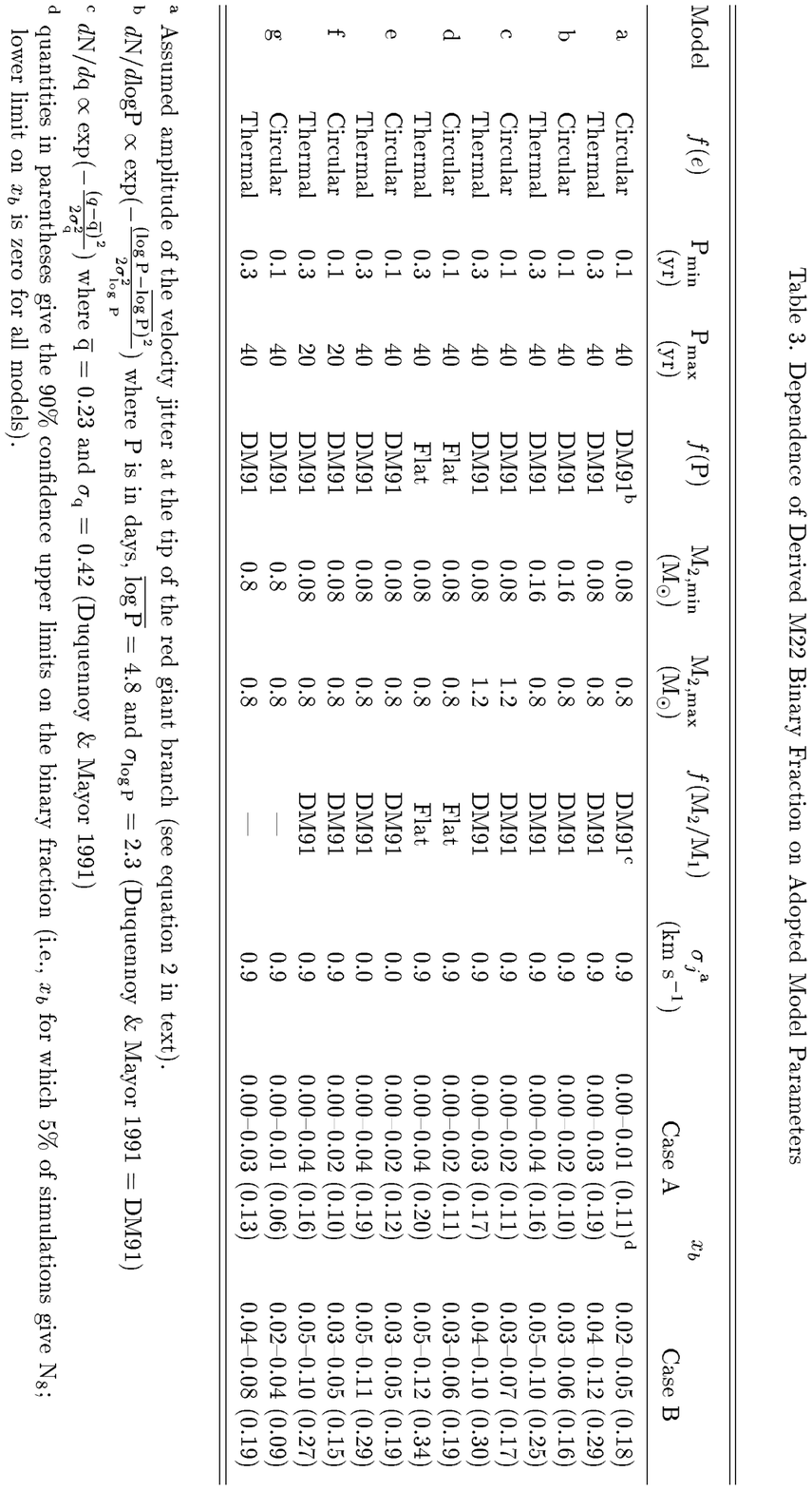}
\vfill\eject
 
\centerline{~}
\includegraphics{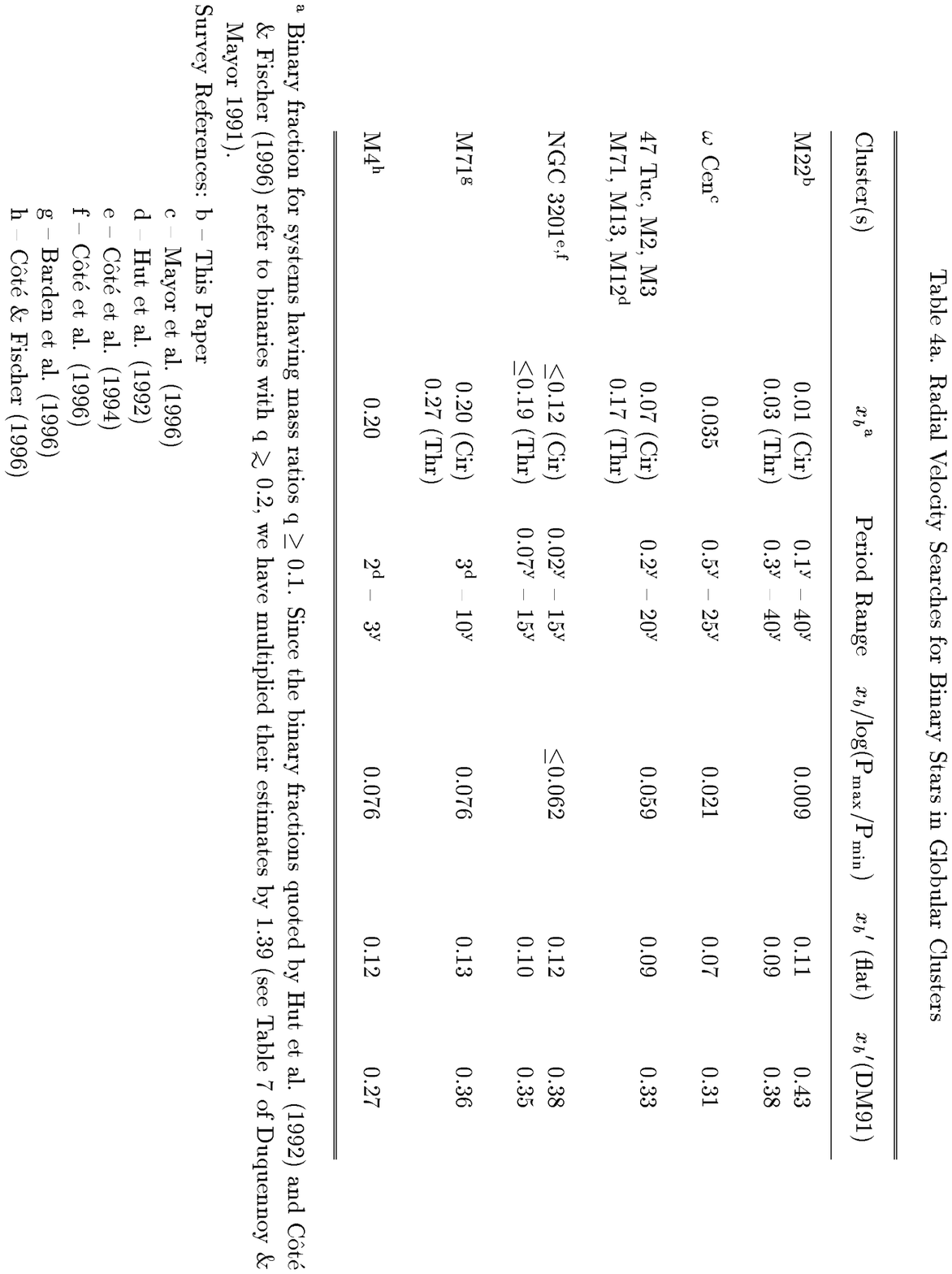}
\vfill\eject
 
\centerline{~}
\includegraphics{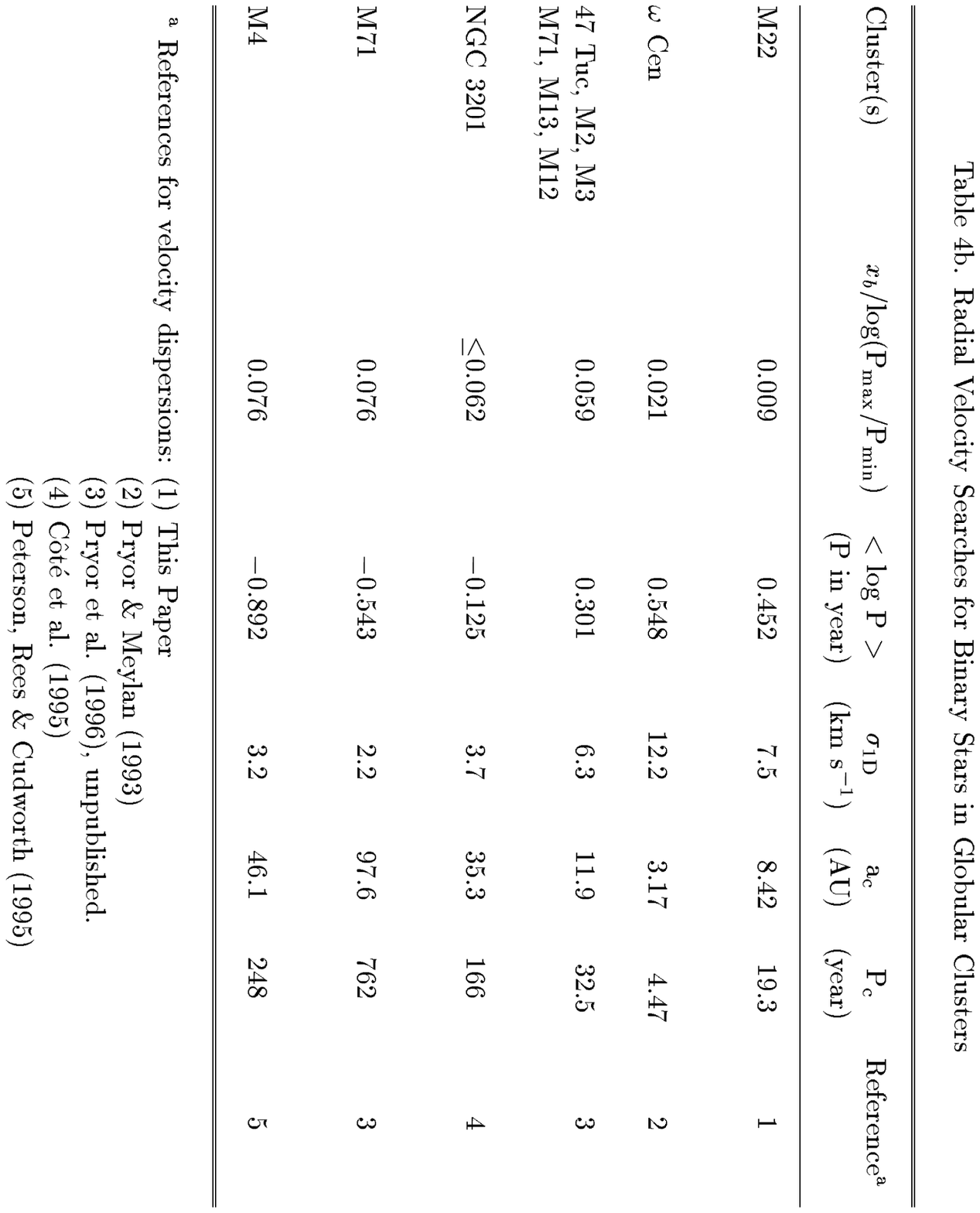}
\vfill\eject

\clearpage
\begin{figure}
\plotone{m22.fig01.ps}
\caption{
V-band finding chart for the six stars not included in previous photometric catalogs.
The image measures 13$\farcm$6$\times$11$\farcm$6. 
}
\end{figure}

\begin{figure}
\plotone{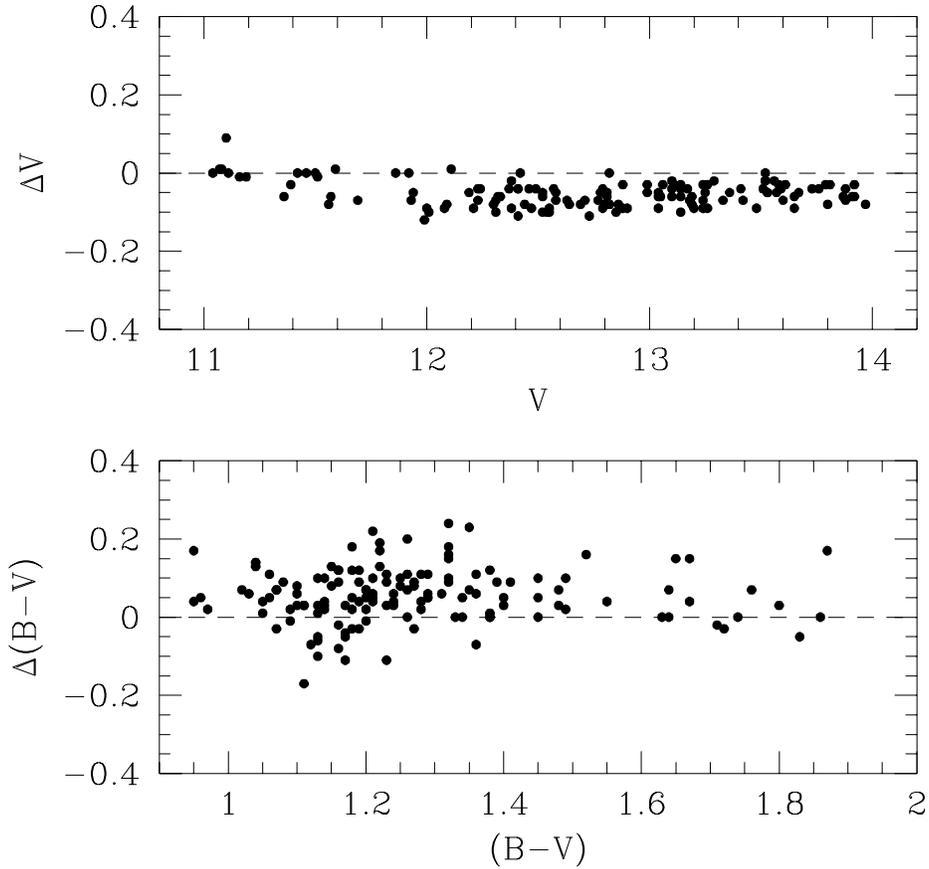}
\caption{
(Upper Panel) Relationship between the V magnitudes reported in this paper and
the photographic values of Peterson \& Cudworth (1994; PC) for the 137 stars in Table 2 which are common to
both studies. The residuals are in the sense $\Delta$V = V$_{\rm PC}$ -- V.
(Lower Panel) Relationship between our (B--V) colors and those of Peterson \& Cudworth (1994)
for the same 137 stars. The residuals are in the sense $\Delta$(B--V) = (B--V)$_{\rm PC}$ -- (B--V).
}
\end{figure}
 
\begin{figure}
\plotone{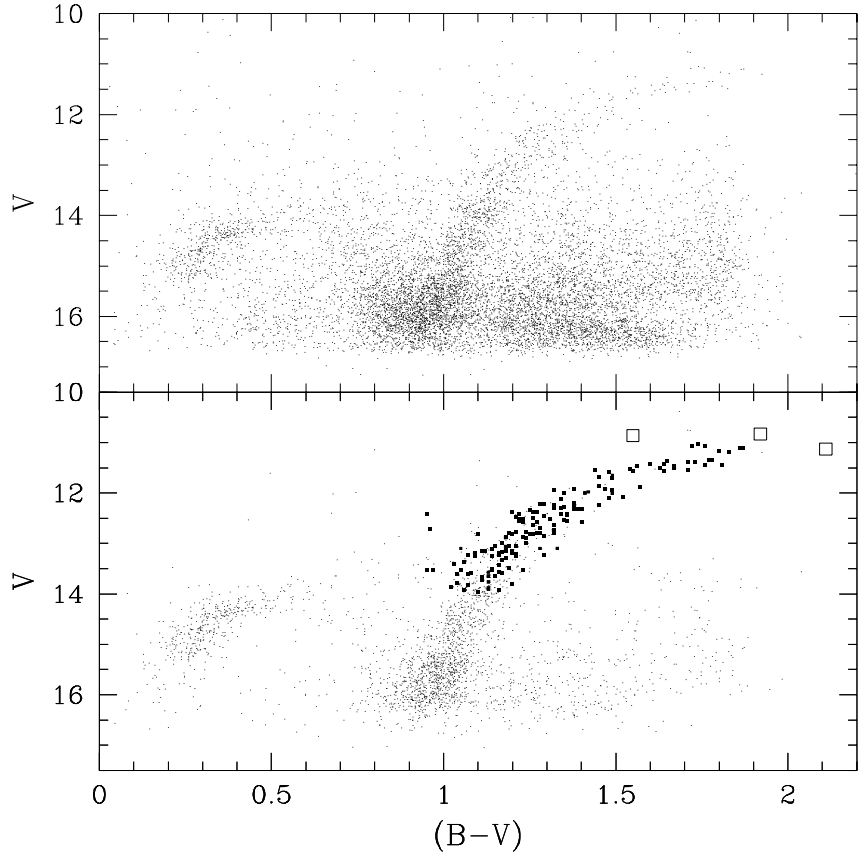}
\caption{
{\fsmall
(Upper Panel) BV color-magnitude diagram for all 9752 stars contained within our
43$\farcm$2$\times$43$\farcm$2 field, based on data obtained with the KPNO 0.9m telescope and T2KA CCD.
(Lower Panel) BV color-magnitude diagram for those 2539 stars within $5r_c$ ($r_c$ = 1$\farcm$42
according to Trager, King \& Djorgovski 1995) of the M22 cluster center (Shawl \& White 1986). 
The stars for which we have multiple radial velocities are indicated by the filled squares.
The large open squares show the three known photometric variables in our survey.
}
}
\end{figure}
 
\begin{figure}
\plotone{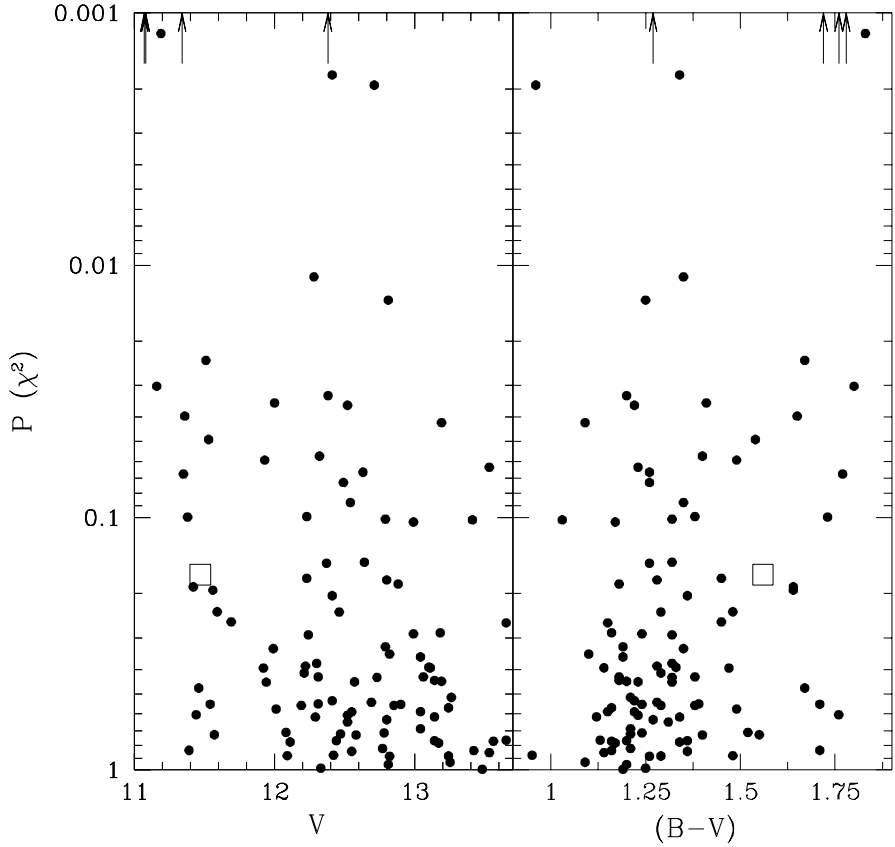}
\caption{
{\fsmall
(Left Panel) Dependence of P($\chi^2$) on V magnitude.
No velocity jitter has been included in computing P($\chi^2$).
The four stars which have probabilities less than 0.001 are indicated by the vertical arrows. 
For V-23 (indicated by the open square), we have discarded the July 1989 CFHT velocity in 
computing P($\chi^2$); P($\chi^2$) $\approx$ 0 otherwise.
(Right Panel) Dependence of P($\chi^2$) on B-V color.
Those stars with the lowest P($\chi^2$) tend to lie near the 
tip of the red giant branch, consistent with previous findings (e.g., Mayor et al. 1984; 
Pryor, Latham \& Hazen 1988) that the magnitude of the velocity jitter is a function of 
luminosity in these evolved stars.
}
}
\end{figure}

\begin{figure}
\plotone{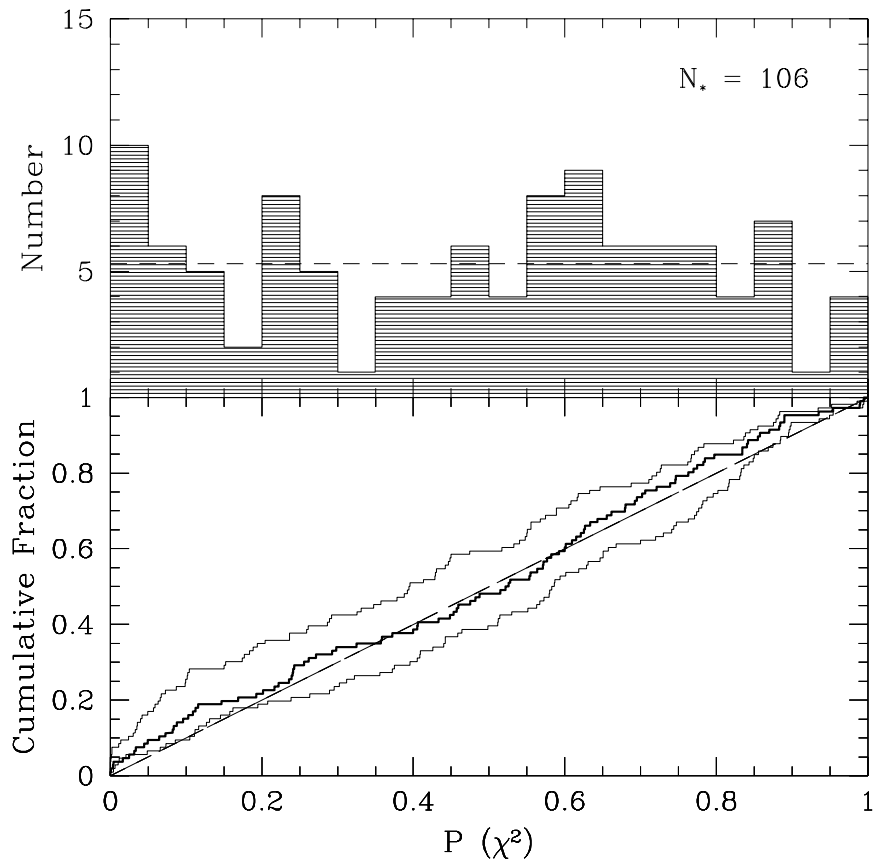}
\vskip-0.5truein
\caption{
{\fsmall
(Upper Panel) Distribution of P($\chi^2$) for the 106 M22 red giants (i.e., three known photometric 
variables excluded) having multiple radial velocity measurements. We have assumed a velocity jitter 
of the form given by equation 3. The radial velocity measured for star V-23 in July 1989 
has been omitted since it is likely to be a misidentification.
A sample of constant-velocity stars is expected to show a flat distribution (dashed line), whereas that 
for a sample of radial velocity variables should be strongly peaked at P($\chi^2$) $\simeq$ 0. 
Apart from a modest peak near zero probability, the observed distribution is consistent with that
expected for a population of constant-velocity stars.
(Lower Panel) Cumulative distribution of P($\chi^2$) for the same sample of M22 giants (heavy line).
The upper and lower lines give the corresponding distributions
assuming no jitter and a luminosity-independent jitter amplitude of 1.5 km s$^{-1}$, respectively.
The dashed diagonal line shows distribution expected for a sample of constant-velocity
stars.
}
}
\end{figure}

\begin{figure}
\plotone{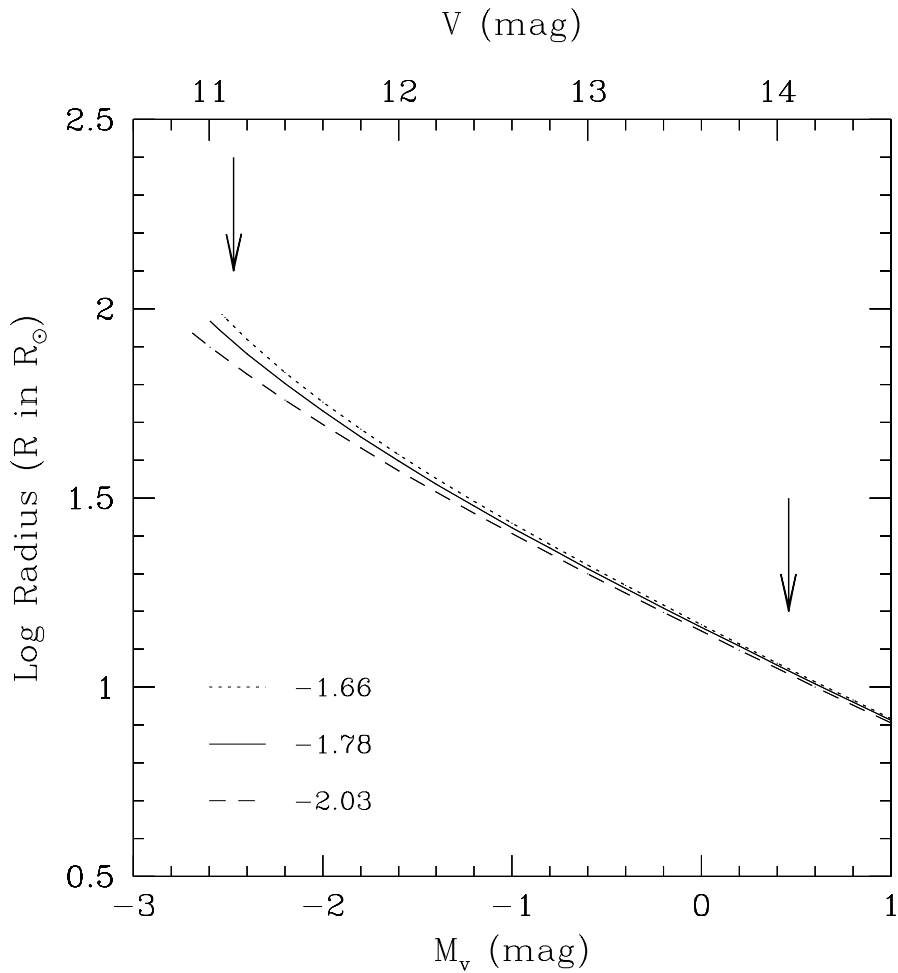}
\caption{
{\fsmall
Relationship between stellar radius, V magnitude (upper axis) and
absolute magnitude M$_{\rm V}$ (lower axis) for M22 red giants according to the 14 Gyr 
isochrones of Bergbusch \& VandenBerg (1992). The three different curves 
have [Fe/H] = --2.03, --1.78 and --1.66, showing the dependence of radius on metallicity. 
The vertical arrows depict the magnitude limits of the survey, which 
correspond to maximum and minimum radii of $\simeq$ 80 and 10R$_{\odot}$, respectively.
In deriving the radii, we have interpolated in V using the [Fe/H] = --1.78 isochrone and 
assumed (m--M)$_{\rm V}$ = 14.2, M$_{\rm V}$(HB) =  0.6, E(B--V) = 0.37 and A$_{\rm V}$ = 3.2E(B--V).
}
}
\end{figure}

\begin{figure}
\plotone{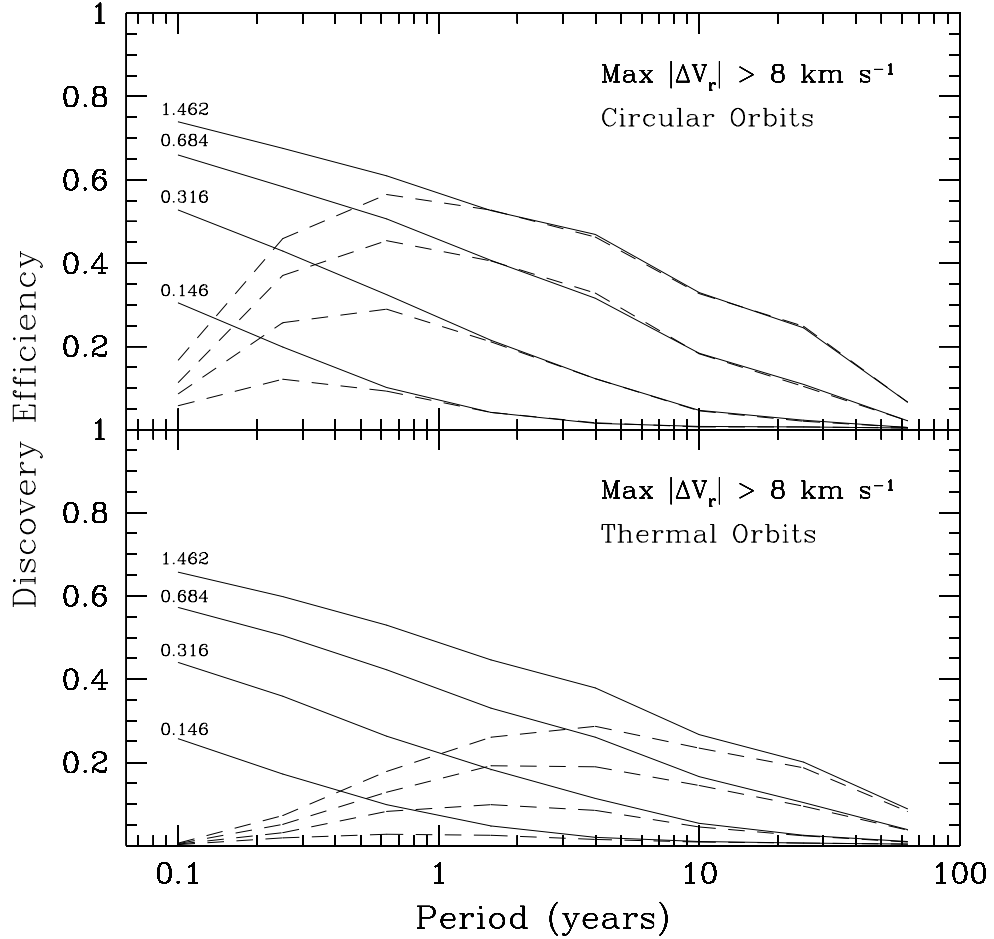}
\caption{
{\fsmall
(Upper Panel) Binary discovery efficiencies based on the data given in Table 2 for the case of
circular orbits. ``Discovered" binaries are those which show a velocity 
variation larger than 8 km s$^{-1}$.
From top to bottom, the four solid lines (labeled by the mean mass ratio in each bin) indicate the 
discovery efficiencies for systems having mass ratios in the intervals: 
(1) 0.00 $\le$ $\log$ q $<$ 0.33;
(2) -0.33 $\le$ $\log$ q $<$ 0.00;
(3) -0.67 $\le$ $\log$ q $<$ -0.33;
and (4) -1.00 $\le$ $\log$ q $<$ -0.67. 
The dashed lines show the discovery efficiencies after taking into account selection effects 
caused by possible mass transfer between the binary components.
(Lower Panel) Same as above, except for a thermal distribution of eccentricities.
}
}
\end{figure}

\begin{figure}
\plotone{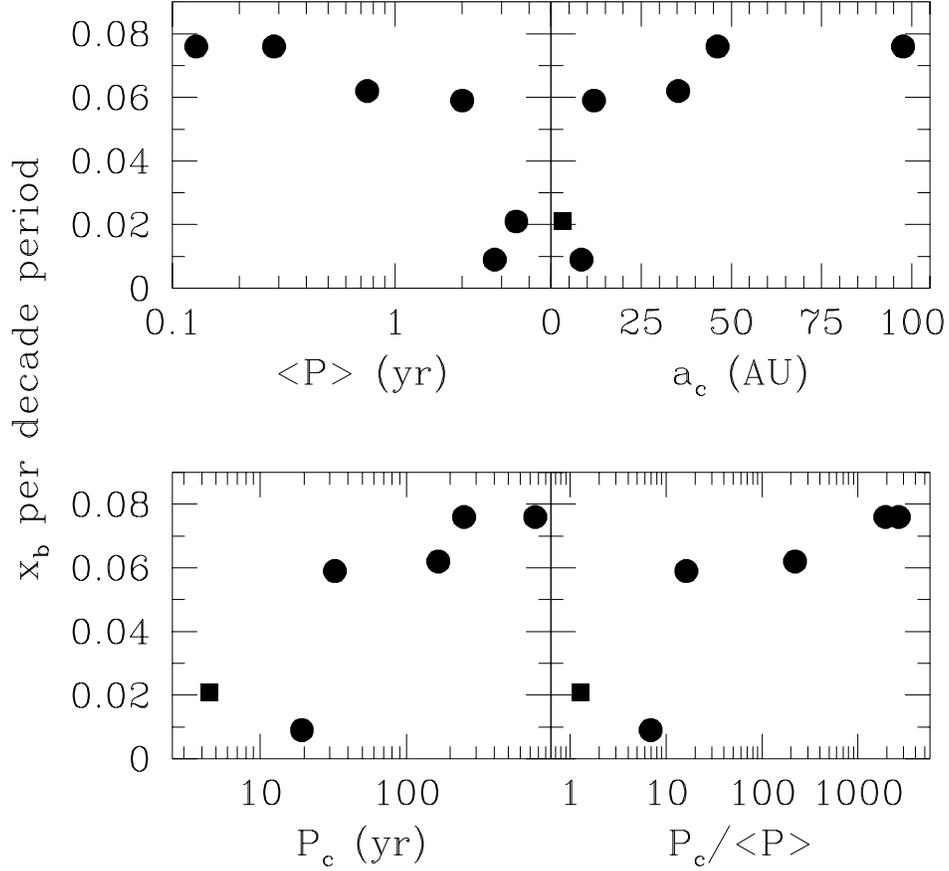}
\caption{
(Upper Left Panel) Binary fraction per decade of period plotted against the logarithmic mean period $<$P$>$ of each
survey (see Table 4b). The binary fraction for NGC 3201 is probably best 
viewed as an upper limit (C\^ot\'e et al. 1994; 1996).
(Upper Right Panel) Binary fraction per decade of period plotted against the critical binary separation a$_{\rm c}$
computed with equation 1 (assuming M$_1$ = M$_2$ = 0.8M$_{\odot}$).
Note that $\omega$ Cen, which is indicated by the closed square, does not satisfy equation 2.
(Lower Left Panel) Binary fraction per decade of period plotted against the critical binary period P$_{\rm c}$ for
a binary consisting of a pair of 0.8M$_{\odot}$ stars.
(Lower Right Panel) Binary fraction per decade of period plotted against P$_{\rm c}$/$<$P$>$. 
{\fsmall
}
}
\end{figure}

% That's all, folks.
%
% The technique of segregating major semantic components of the document
% within "environments" is a very good one, but you as an author have to
% come up with a way of making sure each \begin{whatzit} has a corresponding
% \end{whatzit}.  If you miss one, LaTeX will probably complain a great
% deal during the composition of the document.  Occasionally, you get away
% with it right up to the \end{document}, in which case, you will see
% "\begin{whatzit} ended by \end{document}".


\clearpage

\begin{thebibliography}{}
\bibitem[Anthony-Twarog et al. 1995]{ant95} \reference Anthony-Twarog, B. J., Twarog, B. A., \& Craig, J.
1995, \pasp, 107, 32
\bibitem[Arp \& Melbourne 1959]{arp} \reference Arp, H. C., \& Melbourne, W. G. 1959, \aj, 64, 28
\bibitem[Alcaino et al. 1988]{al88} \reference Alcaino, G., Liller W., \& Alvarado F. 1988, \apj, 330, 569
\bibitem[Barden, Armandroff \& Pryor 1996]{bap96} \reference Barden, S. C., Armandroff, T. E., \&
Pryor, C. 1996, in The Origins, Evolution and Destinies of Binary Stars in Clusters, edited by E. F. Milone, in press
\bibitem[Cohn 1980]{c80} \reference Cohn, H. 1980, \apj, 242, 765
\bibitem[C\^ot\'e et al. 1994]{c94} \reference C\^ot\'e, P., Welch, D.L., Fischer, P., 
Da Costa, G. S., Tamblyn, P., Seitzer, P., \& Irwin, M. J. 1994, \apjs, 90, 83
\bibitem[C\^ot\'e et al. 1995]{c95} \reference C\^ot\'e, P., Welch, D. L., Fischer, P., \& Gebhardt, K. 1995, \apj, 454, 788
\bibitem[C\^ot\'e \& Fischer 1996]{c96} \reference C\^ot\'e, P., \& Fischer, P. 1996, \aj, submitted 
\bibitem[C\^ot\'e et al. 1996]{crfp96} \reference C\^ot\'e, P., Fischer, P., Gebhardt, K., 
Williams, T.B, Pryor, C. 1996, in preparation
\bibitem[Cudworth 1986]{c86} \reference Cudworth, K. M. 1986, \aj, 92, 348
\bibitem[Djorgovski \& King 1986]{dk86} \reference Djorgovski, S. G. \& King, I. R. 1986, \apj, 305, L61
\bibitem[Djorgovski 1993]{d93} \reference Djorgovski, S. G. 1993, 
in The Structure and Dynamics of Globular Clusters, ASP Conference Series, Vol 50, edited by S. G.
Djorgovski and G. Meylan, (ASP, San Francisco), p. 373
\bibitem[Duquennoy \& Mayor 1991]{dm91} \reference Duquennoy, A., \& Mayor, M. 1991, \aap, 248, 485
\bibitem[Fletcher 1982]{f82} \reference Fletcher, J. M., Harris, H. C., McClure, R. D., \& Scarfe, C. D. 1982, PASP, 94, 1017
\bibitem[Gebhardt et al. 1995]{geb95} \reference Gebhardt, K., Pryor, C., Williams, T. B., 
\& Hesser, J. E. 1995, \aj, 110, 1699
\bibitem[Griffin \& Gunn 1974]{grif74} \reference Griffin, R. F., \& Gunn, J. E. 1974, \apj, 191, 545
\bibitem[Gunn \& Griffin 1979]{gunn79} \reference Gunn, J. E., \& Griffin, R. F. 1979, \aj, 84, 752
\bibitem[Hesser \& Harris 1979]{hesser79} \reference Hesser, J. E., \& Harris, G. L. H. 1979, \apj, 234, 513
\bibitem[Hesser et al. 1977]{hesser77} \reference Hesser, J. E., Hartwick, F. D. A., McClure, R. D. 1977, \apjs, 33, 471
\bibitem[Hills 1984]{hills84} \reference Hills, J. G. 1984, \aj, 89, 1811
\bibitem[Heggie 1975]{heg78} \reference Heggie, D. C. 1975, \mnras, 173, 729
\bibitem[Heggie 1980]{heg80} \reference Heggie, D. C. 1980, in Globular Clusters, NATO Adv. Study Institute, ed.
D. Hanes \& B. Madore (Cambridge Univ. Press), p. 281
\bibitem[Hoffer 1983]{hoffer83} \reference Hoffer, J. B. 1983, \aj, 88, 1420
\bibitem[Hut \& Bahcall 1983]{hb83} \reference Hut, P., \& Bahcall, J. N. 1983, \apj, 268, 319
\bibitem[Hut et al. 1992]{hut92} \reference Hut, P., McMillan, S., Goodman, J., Mateo, M., Phinney, E. S., Pryor, C. P.,
Richer, H. B., Verbunt, F., \& Weinberg, M. 1992, \pasp, 104, 981
\bibitem[Lloyd Evans 1975]{le75} \reference Lloyd Evans, T. 1975, \mnras, 171, 647
\bibitem[Lloyd Evans 1978]{le78} \reference Lloyd Evans, T. 1978, \mnras, 182, 293
\bibitem[Lupton, Gunn \& Griffin 1987]{lup87} \reference Lupton, R. H., Gunn, J. E., \& Griffin, R. F. 1987, \aj, 
93, 1114
\bibitem[Mateo 1993]{mateo93} \reference Mateo, M. 1993, in Blue Stragglers, ed. R. Saffer, (San Francisco: ASP), p. 74
\bibitem[Mateo 1996]{mateo96} \reference Mateo, M. 1996, in The Origins, 
Evolution and Destinies of Binary Stars in Clusters, edited by E. F. Milone, in press
\bibitem[Mayor et al. 1984]{may84} \reference Mayor, M., Benz, W., Imbert, M., Martin, N., Prevot, L., Andersen, J., 
Nordstr\"om, B., Ardeberg, A., Lindgren, H., \& Maurice, E. 1984, A\&A, 134, 118
\bibitem[Mayor et al. 1996]{may96} \reference Mayor, M., Duquennoy, A., Udry, S., Andersen, J., \& Nordstr\"om, B. 
1996, in The Origins, Evolution and Destinies of Binary Stars in Clusters, edited by E. F. Milone, in press
\bibitem[McClure \& Norris 1977]{mn77} \reference McClure, R. D., \& Norris, J. 1977, \apj, 216, L101
\bibitem[McClure \& Woodsworth 1990]{mw90} \reference McClure, R. D., \& Woodsworth 1990, \apj, 352, 709
\bibitem[McMillan \& Hut 1994]{mh94} \reference McMillan, S., \& Hut, P. 1994, \apj, 427, 793
\bibitem[Peterson \& Cudworth 1994]{pc94} \reference Peterson, R. C., \& Cudworth, K. M. 1994, ApJ, 420, 612
\bibitem[Peterson, Rees \& Cudworth 1995]{prc95} \reference Peterson, R. C., Rees, R. F., \& Cudworth, K. M. 1995, ApJ, 443, 124
\bibitem[Phinney 1996]{p96} \reference Phinney, E. S. 1996, in The Origins,
Evolution and Destinies of Binary Stars in Clusters, edited by E. F. Milone, in press
\bibitem[Pryor, Latham \& Hazen 1988]{tad88} \reference Pryor, C., Latham, D. W., \& Hazen, M. L. 1988, \aj, 96, 123
\bibitem[Pryor et al. 1989]{tad89} \reference Pryor, C., McClure, R. D., Fletcher, J. M, \& Hesser, J. E. 
1989, \aj, 98, 596
\bibitem[Pryor et al. 1996]{tad96} \reference Pryor, C., Fletcher, J. M., Hesser, J. E., McClure, R. D., Stetson, P. B., 
Richer, H. B., Fahlman, G. G., Ibata, R. A., Ivanans, N. C., Mandushev, G., Bell, R. A., 
Bolte, M., Bond, H. B., Harris, W. E., VandenBerg, D. A., \& Wood, M. A. 1996, In Dynamical Evolution of Star 
Clusters -- Confrontation of Theory and Observations", IAU Symp 174, eds J. Makino \& P. Hut, in press.
\bibitem[Pryor \& Meylan 1993]{pm93} \reference Pryor, C., \& Meylan, G. 1993, in The Structure and Dynamics 
of Globular Clusters, edited by S.G. Djorgovski \& G. Meylan, (ASP, San Francisco), p. 357
\bibitem[Sawyer-Hogg 1973]{sh73} \reference Sawyer-Hogg, H. 1973, Publ. David Dunlap Obs., 3, 6.
\bibitem[Shawl \& White 1986]{sw86} \reference Shawl, S. J., and White, R. E. 1986, \aj, 91, 312
\bibitem[Sigurdsson \& Phinney 1993]{sp93} \reference Sigurdsson, S., \& Phinney, E. S. 1993, \apj, 415, 631
\bibitem[Stetson, Davis, \& Crabtree 1990]{s90} \reference Stetson, P. B., Davis, L. E., \& Crabtree, D. R. 1990, in
CCDs in Astronomy, ASP Conference Series, Vol.25, edited by G. H. Jacoby (ASP, San Francisco), p. 297
\bibitem[Stetson 1994]{s94} \reference Stetson, P. B. 1994, \pasp, 106, 250
\bibitem[Trager et al. 1995]{trag95} \reference Trager, S., King, I. R., \& Djorgovski, S. G. 1995, \aj, 109, 218
\bibitem[Vanture \& Wallerstein 1992]{vw92} \reference Vanture, A. D., \& Wallerstein, G. 1992, \pasp, 104, 888
\bibitem[Yan \& Mateo 1994]{ym94} \reference Yan, L., \& Mateo, M. 1994, \aj, 108, 1810
\bibitem[Zinn \& Searle 1976]{zs76} \reference Zinn, R., \& Searle, L. 1976, \apj, 209, 734
\end{thebibliography}
\end{document}